\documentclass[preprint,preprintnumbers,amsmath,amssymb]{revtex4-1}
\usepackage[dvipsnames,table]{xcolor}
 \usepackage[colorlinks=false]{hyperref}
 \usepackage{lipsum}
 \usepackage{graphicx}
\usepackage{latexsym}
\usepackage{amsmath}
\usepackage{amsthm}
\usepackage{amssymb}
\usepackage{epstopdf} 
\usepackage{enumerate}
\usepackage{setspace} % controllabel line spacing
\usepackage{dcolumn}% Align table columns on decimal point
\usepackage{bm}% bold math
\usepackage{setspace} % controllabel line spacing
\usepackage{slashed}
\usepackage{color}
\usepackage{youngtab}
\usepackage{tikz}
\usepackage{tikz-3dplot}
\usepackage{braket}
\usetikzlibrary{shapes,snakes,arrows,chains,matrix,positioning,scopes,calc}
\usetikzlibrary{decorations.markings}

\usepackage[tabbotcap]{subfigure}
\tikzset{/pgf/decoration/.cd,
    number of sines/.initial=10,
    angle step/.initial=20,
}
\newdimen\tmpdimen
\pgfdeclaredecoration{complete sines}{initial}
{
    \state{initial}[
        width=+0pt,
        next state=move,
        persistent precomputation={
            \pgfmathparse{\pgfkeysvalueof{/pgf/decoration/angle step}}%
            \let\anglestep=\pgfmathresult%
            \let\currentangle=\pgfmathresult%
            \pgfmathsetlengthmacro{\pointsperanglestep}%
                {(\pgfdecoratedremainingdistance/\pgfkeysvalueof{/pgf/decoration/number of sines})/360*\anglestep}%
        }] {}
    \state{move}[width=+\pointsperanglestep, next state=draw]{
        \pgfpathmoveto{\pgfpointorigin}
    }
    \state{draw}[width=+\pointsperanglestep, switch if less than=1.25*\pointsperanglestep to final, % <- bit of a hack
        persistent postcomputation={
        \pgfmathparse{mod(\currentangle+\anglestep, 360)}%
        \let\currentangle=\pgfmathresult%
    }]{%
        \pgfmathsin{+\currentangle}%
        \tmpdimen=\pgfdecorationsegmentamplitude%
        \tmpdimen=\pgfmathresult\tmpdimen%
        \divide\tmpdimen by2\relax%
        \pgfpathlineto{\pgfqpoint{0pt}{\tmpdimen}}%
    }
    \state{final}{
        \ifdim\pgfdecoratedremainingdistance>0pt\relax
            \pgfpathlineto{\pgfpointdecoratedpathlast}
        \fi
   }
}

 \allowdisplaybreaks
\begin{document}
\title{  Quantum Criticalities with Infinite Anisotropy 
\\ in Topological Phase Transitions  between Dirac and Weyl Semi-metals  
}
 
\author{SangEun Han$^1$}
\author{Gil Young Cho$^{2,3}$}
\author{Eun-Gook Moon$^1$}
\thanks{egmoon@kaist.ac.kr}
\affiliation{$^1$Department of Physics, KAIST, Daejeon 34141, Korea}
\affiliation{$^2$School of Physics, Korea Institute for Advanced Study, Seoul 02455, Korea}
\affiliation{$^{3}$Department of Physics, Pohang University of Science and Technology (POSTECH), Pohang 37673, Republic of Korea}

\date{\today}   
\begin{abstract}
We study quantum phase transitions (QPTs) associated with splitting nodal Fermi points, motivated by topological phase transitions between Dirac and Weyl semi-metals. A Dirac point in Dirac semi-metals may be split into two Weyl points by breaking a lattice symmetry or time-reversal symmetry, and the Lifshitz transition is commonly used to describe the phase transitions. Here, we show that the Lifshitz description is fundamentally incorrect in QPTs with splitting nodal Fermi points. We argue that correlations between fermions, order parameter, and the long-range Coulomb interaction { must} be incorporated from the beginning. 
One of the most striking correlation effects we find is {\it infinite anisotropy} of physical quantities, which cannot appear in a Lifshitz transition.  
By using the standard renormalization group (RG) method, two types of infinitely anisotropic quantum criticalities are found in three spatial dimensions varying with the number of the Dirac points ($N_f$). 
For  $N_f = 1$, the ratio of the fermion velocity to the velocity of order parameter excitations becomes universal  ($1+\sqrt{2}$) along the Dirac point splitting direction . For $N_f >1$, we find that fermions are parametrically faster than order parameter excitations in all directions.
Our RG analysis is fully controlled by the fact that order parameter and fermion fluctuations are at the upper critical dimension, and thus our stable fixed points demonstrate the presence of weakly coupled quantum criticalities with infinite anisotropy. 
%In other words, correlation effects {\it must} be considered from the beginning in topological phase transitions between Dirac and Weyl semi-metals.  
\end{abstract}

\maketitle

\section{Introduction}

Recent advances in topological insulators and semi-metals deepen our understanding in phases and their transitions \cite{RevModPhys.82.3045,RevModPhys.83.1057,RevModPhys.90.015001}.
Incorporating lattice symmetries, insulating and semi-metallic phases are largely classified  in non / weakly interacting  systems, for example topological crystalline insulators \cite{PhysRevLett.106.106802,doi:10.1146/annurev-conmatphys-031214-014501,Hsieh2012} and Dirac line nodal semi-metals \cite{PhysRevLett.115.036806, PhysRevB.93.035138, 1674-1056-25-11-117106}. 
If topological invariants of such phases are protected by lattice symmetries, breaking the protecting symmetries may induce topological phase transitions. Topology and symmetry become intrinsically tied in such quantum phase transitions, and novel quantum criticalities may emerge out of the interplay between topology and symmetry. 

Ignoring order parameter fluctuations, a topological phase transition is often described by the Lifshitz transition, band-structure changing transition \cite{doi:10.1063/1.4974185}. Especially, Dirac / Weyl systems in three spatial dimensions (3d)  such as in BiZnSiO$_{4}$ and Cd$_3$As$_2$ \cite{PhysRevLett.112.036403,Liu2014,Liu864,PhysRevB.95.035102,PhysRevB.83.205101,PhysRevB.85.045124} have marginally correlated excitations \cite{PhysRevLett.108.046602}  in sharp contrast to strongly correlated systems where fermionic excitations are strongly coupled \cite{PhysRevLett.85.4940, PhysRevB.78.064512, PhysRevB.80.165102, PhysRevB.82.075127,  PhysRevB.82.075128, PhysRevB.82.045121,PhysRevLett.111.206401, PhysRevB.89.201110, PhysRevLett.113.106401, PhysRevX.4.041027,  PhysRevB.92.045117, PhysRevB.92.035141,PhysRevB.94.195135,  Cho2016,  PhysRevLett.116.076803, PhysRevB.95.075149,   PhysRevB.95.094502, hong2017, PhysRevB.96.155112, PhysRevX.7.021010,    doi:10.1146/annurev-conmatphys-031016-025531}. Thus, their transitions are believed to be described by the Lifshitz transition.
Indeed, a certain class of topological phase transitions is well described by the Lifshitz transition of the Dirac fermions up to logarithmic corrections. For example, topological phase transitions between nodal and nodeless superconductors, are described by the Gross-Neveu-Yukawa theory whose coupling constants are marginally irrelevant in 3d \cite{ZINNJUSTIN1991105,sachdev2011,2018arXiv180205727H, PhysRevB.87.205138,Yang2014, PhysRevB.93.241113,Roy2016,  PhysRevB.95.201102}. 
Then, it is natural to ask whether the Lifshitz transition description always works in 3d Dirac systems. In this paper, we concretely show that the Lifshitz transition even fails in a class of topological phase transitions of 3d Dirac systems, namely QPTS  associated with splitting nodal Fermi points. Correlation effects from order parameter fluctuations and the long-range Coulomb interaction {\it must} be incorporated from the beginning. 
 
\begin{figure}[t]
\centering
\includegraphics[width=\linewidth]{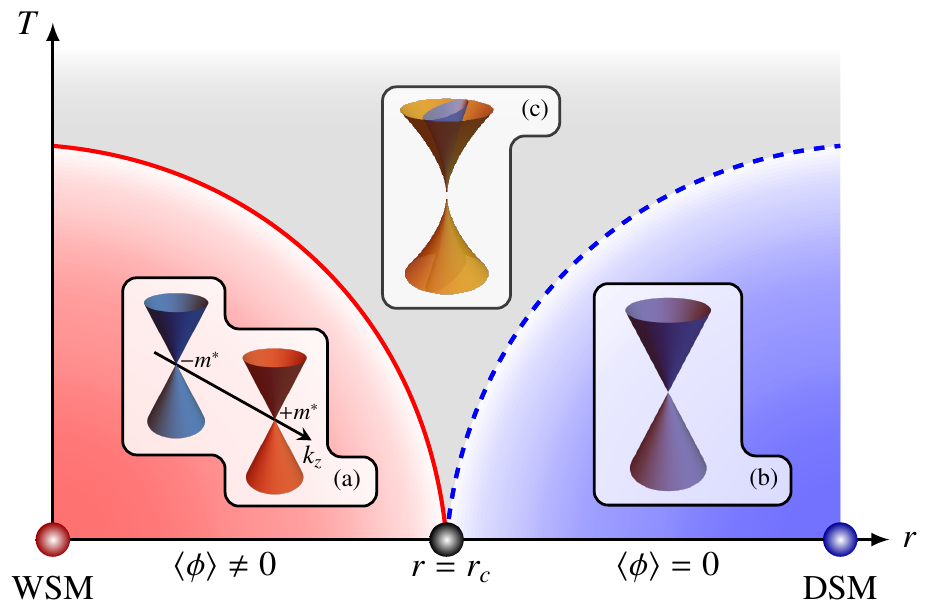}
\caption{Topological phase transition between WSM and DSM. In a symmetric phase ($r>r_{c}$), a Dirac point is illustrated, and in a symmetric broken phase,  two Weyl points with different chirality (black and white points) are illustrated. (a), (b), and (c) represent the fermion energy dispersion relations of WSM, DSM, and quantum critical point, respectively. }\label{fig:dirac_weyl}
\end{figure}

We focus on a minimal model of a Dirac semi-metal (DSM). Dirac points in the Brillouin zone, where valence and conduction bands touch linearly with four degenerate states, are protected by {a set of lattice symmetries and time-reversal symmetry} \cite{PhysRevB.85.195320, PhysRevB.88.125427,RevModPhys.90.015001,PhysRevLett.112.036403,Liu864,Liu2014,PhysRevB.95.035102}. Breaking the protecting lattice symmetry may induce a Weyl semi-metal (WSM) where two Weyl points with non-zero Berry flux around the Weyl points appear as illustrated in Fig.\ref{fig:dirac_weyl}. 
We emphasize that the Dirac point splitting indicates that an order parameter of the protecting symmetry is coupled to fermions {\it non-relativistically}.
Since a pair of the Weyl points has the opposite signs of the Berry flux, it is obvious that the Berry flux around the Dirac point vanishes. 
The direction which connects the two Weyl points is special, and we set it as a $z$ direction in this paper.
Notice that the Dirac and Weyl points are not generically located at the zero energy (chemical potential) along the quantum phase transitions unless additional symmetries protect, for example, such as particle-hole or sub-lattice symmetries.  We ignore such chemical potential issues in this paper to investigate intrinsic properties of topological phase transitions with nodal point splitting. Moreover, recent advances in material engineering suggest possibilities of semi-metals without electron-hole pockets as in BiZnSiO$_{4}$ \cite{PhysRevLett.112.036403}.

%Dirac to Weyl

Three types of low energy excitations exist around phase transitions between DSM and WSM; fermions, the long-range Coulomb interaction, and order parameter fluctuations. 
We investigate their interplay and show the Lifshitz transition is intrinsically insufficient to understand the topological phase transitions associated with splitting nodal points. We employ the standard RG analysis with the momentum-shell scheme with all the excitations \cite{sachdev2011} and obtain stable fixed points indicating continuous QPTs. 
The strength of the fine structure constants of electric charge and Yukawa coupling become marginally irrelevant similar to relativistic quantum field theories with Dirac fermions and boson excitations at the upper critical dimension. Yet, we find striking characteristics of the topological phase transitions emerged from the interplay between the low energy degrees of freedom. For example, the anisotropy of the velocities of excitations becomes {\it universal} in sharp contrast to the ones of the Lifshitz theory where all velocities are arbitrary. Moreover, the universal number is not a unity which demonstrates that our critical theories are described by non-relativistic quantum field theories. Most of the velocity ratios are infinite demonstrating infinitely anisotropic quantum criticalities. 
To demonstrate the infinite anisotropy manifestly,  we keep all spatial anisotropies of the excitations in our calculations. In other words, our analysis is non-perturbative in the anisotropies while it is perturbative in the coupling constants $(\alpha_e, \alpha_g)$.

Infinitely anisotropic quantum criticalities have been suggested in strongly coupled quantum critical points. Huh and Sachdev first show its possibility in nematic transitions of d-wave superconductors in 2D \cite{PhysRevB.78.064512}, and Savary {\it et. al.} show a similar criticality in the Luttinger semi-metals with the long-range Coulomb interaction \cite{PhysRevX.4.041027}. 
In both cases, the universality classes are strongly coupled in a sense that their order parameters receive large anomalous dimensions. 
The calculations are controlled by a fermion flavor number and infinite anisotropy. 
We emphasize that the infinite anisotropic quantum criticalities we find in this paper are {\it weakly} coupled in a sense that an anomalous dimension of order parameters is zero.  
We find the two universality classes varying with the number of the Dirac points $N_f$. For $N_f >1$, all velocity ratios become infinite, and for $N_f=1$, we find one more universal velocity ratio ($1+\sqrt{2}$) as shown below.

We will also generalize our methods to approach strongly coupled regimes by considering a non-zero finite fine structure constant, which may induce non-Fermi liquid behaviors. Even though our calculations lose reliability a bit,  we find intriguing effects on infinitely anisotropic quantum criticalities at strong coupling regime, which may be naturally interpolated to previously studied strongly coupled QPTs with infinite anisotropy.

The structure of this paper is as follows. In section II, we introduce a model Hamiltonian for topological phase transitions between Dirac and Weyl semi-metals. 
The full action with all low energy excitations is explicitly written, and dimensionless coupling constants are listed, which are useful for the RG calculations.  Detailed RG calculations are presented in section III and we analyze RG equations for three cases; 1) the case without the Coulomb interaction, 2) the case with the Coulomb interaction, and 3) the case with non-Fermi liquids. Section IV presents discussion and conclusion.

\section{Model}

\begin{table}
\begin{tabular}{|>{$}c<{$} |c |}
\hline
\text{Parameters}&\begin{tabular}{@{}r@{}}Descriptions\end{tabular}\\
\hline\hline
v_{z}& $z$-directional Fermion velocity \\\hline
v_{\perp}& $x$,$y$- directional Fermion velocity  \\\hline
u_{z}& $z$-directional order parameter velocity\\\hline
u_{\perp}&\begin{tabular}{@{}l@{}}  $x$,$y$- directional order parameter velocity \end{tabular} \\\hline
e&\begin{tabular}{@{}l@{}}  electric charge \end{tabular}\\\hline
g&\begin{tabular}{@{}l@{}} Yukawa coupling constant \end{tabular}\\\hline
\lambda& quartic coupling constant of order parameter \\\hline \hline
\begin{tabular}{@{}>{$}c<{$}@{}}\gamma_{f}  \equiv{v_{z}}/{v_{\perp}}\end{tabular}&\begin{tabular}{@{}l@{}} fermion velocity ratio\end{tabular}\\\hline
\begin{tabular}{@{}>{$}c<{$}@{}}\gamma_{b} \equiv{u_{z}}/{u_{\perp}}\end{tabular}&\begin{tabular}{@{}l@{}} order parameter velocity ratio\end{tabular}\\\hline
\gamma_{c}&  Coulomb interaction anisotropy constant \\\hline
\begin{tabular}{@{}>{$}c<{$}@{}}R_{\perp} \equiv u_{\perp}/v_{\perp}\end{tabular}& \begin{tabular}{@{}l@{}}fermion-order parameter velocity ratio \\in the  $x,y$ directions\end{tabular}\\\hline
\begin{tabular}{@{}>{$}c<{$}@{}}R_{z} \equiv u_{z}/v_{z}\end{tabular}& \begin{tabular}{@{}l@{}} fermion-order parameter velocity ratio \\in the  $z$ direction\end{tabular}\\\hline
\begin{tabular}{@{}>{$}c<{$}@{}}R_{fc} \equiv \gamma_{f}/\gamma_{c}\end{tabular}&\begin{tabular}{@{}l@{}}fermi-Coulomb anisotropy ratio \end{tabular}\\\hline
\begin{tabular}{@{}>{$}c<{$}@{}}\alpha_{e} \equiv{\frac{e^{2}}{4\pi v_{z}}}\end{tabular}&\begin{tabular}{@{}l@{}}fine structure constant with electric charge \end{tabular}    \\\hline
\begin{tabular}{@{}>{$}c<{$}@{}}\alpha_{g}\equiv{\frac{g^{2}}{4\pi v_{z}}}\end{tabular}&\begin{tabular}{@{}l@{}}fine structure constant with Yukawa coupling\end{tabular}    \\\hline
\begin{tabular}{@{}>{$}c<{$}@{}}\tilde{\lambda}\equiv{\frac{\lambda}{\gamma_{b}}}\end{tabular}&\begin{tabular}{@{}l@{}}redefined  quartic coupling constant\end{tabular}\\\hline
\end{tabular}
\caption{List of parameters for physical quantities and dimensionless ratios. }
\end{table}

We start with a low energy Hamiltonian of DSM,  
\begin{align}
H_0=\sum_{\bm{k}}\Psi_{\bm{k}}^{\dagger}\mathcal{H}_0(\bm{k})\Psi_{\bm{k}}=\sum_{\bm{k}}\Psi_{\bm{k}}^{\dagger}(\vec{d}(\bm{k})\cdot\vec{\Gamma})\Psi_{\bm{k}},
\end{align}
where $\Psi$ is $4N_{f}$-component spinor. The functions ($\vec{d}(\bm{k})$) of momentum ($\bm{k}$) are for an energy dispersion relation, and the matrices ($\vec{\Gamma}$) are for the Clifford algebra, $\{\Gamma_{i},\Gamma_{j}\}=2\delta_{ij}$. Generically, the Dirac fermion has a linear dispersion relation with $d_{i}(\bm{k})=v_{i}k_{i}$ ($i=x,y,z$) unless additional symmetries are present. One representation for the Clifford algebra is $\Gamma_{i}=\tau_{z}\otimes\sigma_{i}\otimes I_{N_{f}}$ where $\tau_{i}$ and $\sigma_{i}$ are Pauli matrices and $I_{N_{f}}$ is the $N_{f}\times N_{f}$ identity matrix. So $\Gamma$'s are $4N_{f}\times4N_{f}$ matrices and we can easily confirm that $\Gamma_{i}$ satisfies the Clifford algebra, $\{\Gamma_{i},\Gamma_{j}\}=2\delta_{ij}I_{4N_{f}}$. The energy dispersion is  $E(\bm{k})=\pm\sqrt{v_{x}^{2}k_{x}^{2}+v_{y}^{2}k_{y}^{2}+v_{z}^{2}k_{z}^{2}}$, and  $4N_f$ states  have zero-energy at the origin ($k=0$).  

The presence of the four degenerate states at the Dirac point is guaranteed by lattice symmetry protection.   
By breaking a protecting lattice symmetry, a Dirac point becomes either gapped or split into two Weyl points generically. For example, 
breaking time-reversal symmetry in distorted spinel structure with one Dirac point at $K$ point split the Dirac point into two Weyl points with non-zero Berry flux.  
Mathematical incorporation of splitting a Dirac point into two Weyl points is straightforward. 
We choose one operator $M$, which commutes with only one of $\Gamma_{i}'s$ and anticommute with the others, which may be achieved by $M\equiv\tau_{0}\sigma_{z}$. Adding $g\phi(\Psi^{\dagger}M\Psi)$ to $H_0$, it is obvious that the two Weyl points at $(0,0,\pm m^{*})$ appear with $m^*= g|\phi|$. 
In a symmetric phase ($|m^*|=0$), only one Dirac point with four degenerate states exists while two Weyl points with two degenerate states in a symmetry broken phase ($|m^*| \neq 0$).

To investigate correlation effects, we employ a model action,
\begin{align}
\notag\mathcal{S}=&\int_{x,\tau}\left[ \Psi ^{\dagger}(\partial_{\tau}+\mathcal{H}_0(-i\nabla))\Psi  \right]\\
\notag&+\int_{x,\tau}\frac{1}{2}\left[ (\partial_{x}\varphi)^{2}+(\partial_{y}\varphi)^{2}+\gamma_{c}^{2}(\partial_{z}\varphi)^{2} \right]\\
\notag&+\int_{x,\tau}\frac{1}{2}\left[ \frac{(\partial_{\tau}\phi)^{2}}{u_{\perp}^{2}}+(\partial_{\perp}\phi)^{2}+\left(\frac{u_{z}}{u_{\perp}}\right)^{2}(\partial_{z}\phi)^{2}+\frac{r}{2u_{\perp}^{2}}\phi^{2}\right]\\
&+\int_{x,\tau}\left[ \frac{1}{4!}\frac{\lambda}{u_{\perp}}\phi^{4}+ie\varphi(\Psi ^{\dagger}\Psi)+g\phi(\Psi ^{\dagger}M\Psi) \right],\label{eq:action}
\end{align}
where the short-handed notation $\int_{x,\tau}\equiv\int d^{3}xd\tau$ is {used}. The instantaneous long range Coulomb interaction is described by $\varphi$  (electric potential), and $\phi$ is for an order parameter. The symbols ($u,v$) are for velocities of the order parameter and fermion velocities, respectively, and their subscripts are for spatial directions. For simplicity, we assume that $v_{x}/v_{y}=u_{x}/u_{y}=1$ setting $v_{x}=v_{y}=v_{\perp}$ and $u_{x}=u_{y}=u_{\perp}$. Its generalization to a general case is straightforward, and  the two ratios ($v_{x}/v_{y}$, $u_{x}/u_{y}$) become the same at a fixed point, which may give an additional overall factor, $v_{y}/v_{x}$ to the fixed point of $v_x=v_y$ (See Appendix \ref{app:xyanisotropy}). The order parameter fluctuation term $(\partial_{\perp}\phi)^{2}$ is equal to $(\partial_{\perp}\phi)^{2}\equiv(\partial_{x}\phi)^{2}+(\partial_{y}\phi)^{2}$. The anisotropy of the Coulomb interaction is represented by $\gamma_{c}$.

Remark that our RG analysis is perturbative in $e, g$ but it is non-perturbative in $v,u$. Thus, we introduce all the parameters of spatial anisotropies and keep them along our analysis. For future convenience, we define the dimensionless coupling constants,
\begin{align*}
 R_{\perp}&\equiv\frac{u_{\perp}}{v_{\perp}},& R_{z}\equiv&\frac{u_{z}}{v_{z}},& R_{fc}\equiv&\frac{v_{z}/v_{\perp}}{\gamma_{c}},\\
\alpha_{g}&\equiv\frac{g^{2}}{4\pi v_{z}},& \alpha_{e}\equiv&\frac{e^{2}}{4\pi v_{z}},&\tilde{\lambda}\equiv&\frac{\lambda}{u_{z}/u_{\perp}}.
\end{align*}
All the parameters and dimensionless coupling constants are summarized in Table I.

Before going further, we remark that recent studies on Coulomb interaction effects on DSM / WSM report possibilities of the non-Fermi liquid phase at the strong coupling limit by using the self-consistent Schwinger-Dyson equation \cite{PhysRevB.96.081104}. The non-Fermi liquid phase may be understood as a stable fixed point with $\left.\frac{d\alpha_{e}}{d\ell}\right|_{\alpha_{e}=\alpha_{e}^{*}}=0$ and $\alpha_{e}^{*}\neq0$, and we investigate how the non-Fermi liquid behaviors affect our quantum criticalities.

\section{Renormalization Group analysis}

The standard RG procedure is used in our analysis. 
For simplicity, we set $r=0$ since we focus on quantum critical point in this work and adopt the Wilsonian momentum-shell procedure. After integrating out frequencies, the ultra-violet (UV) and infra-red (IR) cutoffs of momentums are introduced and we extract information about UV and IR divergences. Later, we check our results are independent of the choice of the cut-off axis (see Appendix \ref{sm:rgscheme}). 
Our results show that at the one-loop order, all integrations only contain logarithmic divergences, so our calculation is fully controlled.

From the Eqn 2, we find that the Green's functions of all the excitations are 
\begin{align}
G_{f,0}(\omega,\bm{k})=&\frac{1}{-i\omega+\mathcal{H}_{0}(\bm{k})},\\
G_{\varphi,0}(\omega,\bm{k})=&\frac{1}{k_{\perp}^{2}+\gamma_{c}^{2}k_{z}^{2}},\\
G_{\phi,0}(\omega,\bm{k})=&\frac{1}{\omega^{2}/u_{\perp}^{2}+k_{\perp}^{2}+(u_{z}/u_{\perp})^{2}k_{z}^{2}},%,
\end{align}
where the subscripts ($f, \varphi, \phi$) are for fermions, Coulomb interaction, and order parameter, respectively. 
All the Feynman diagrams at the leading order are illustrated in Fig 2.

%\onecolumngrid

\begin{figure}[h]
\centering
\subfigure[]{
\includegraphics{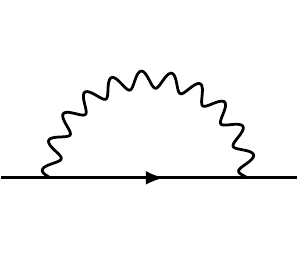}\label{fig1:a}}
\subfigure[]{
\includegraphics{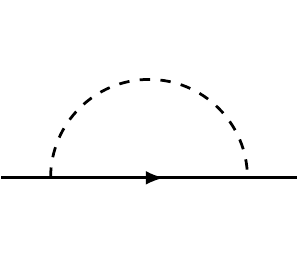}\label{fig1:b}}
\subfigure[]{
\includegraphics{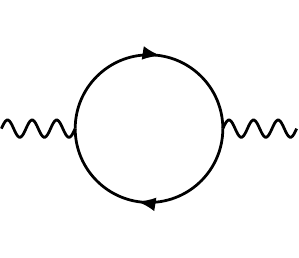}\label{fig1:c}}
\subfigure[]{
\includegraphics{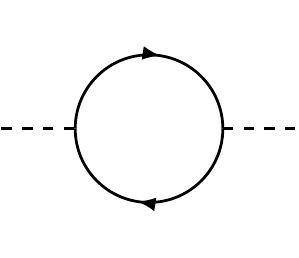}\label{fig1:d}}
\subfigure[]{
\includegraphics{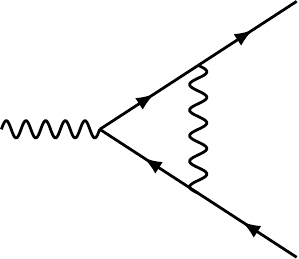}\label{fig1:e}}
\subfigure[]{
\includegraphics{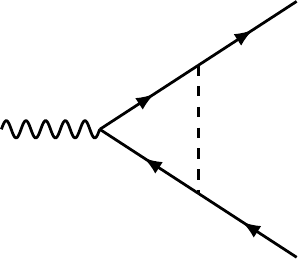}\label{fig1:f}}
\subfigure[]{
\includegraphics{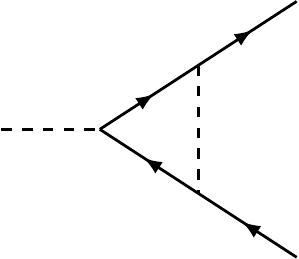}\label{fig1:g}}
\subfigure[]{
\includegraphics{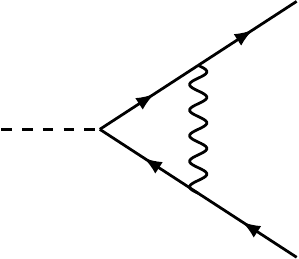}\label{fig1:h}}
\subfigure[]{
\includegraphics{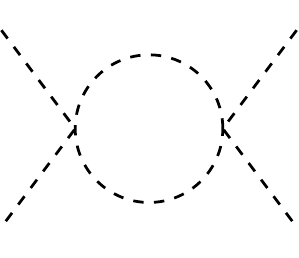}\label{fig1:i}}
\subfigure[]{
\includegraphics{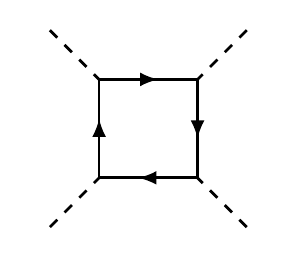}\label{fig1:j}}
\caption{Feynman diagrams at the one-loop order. The line with arrowhead, dashed line, and wavy line stand for the fermion, the order parameter, and the Coulomb interaction, respectively.}\label{fig:feynmann}
\end{figure}
%\twocolumngrid

The fermion self energy can be obtained by evaluating the diagrams of Fig.\ref{fig1:a}, \ref{fig1:b},
\begin{align*}
&\Sigma_{b}(\omega,\bm{k}) =g_{b}^{2}\int_{\Omega,q}M_{b}G_{f,0}(\omega+\Omega,\bm{k}+\bm{q})M_{b}G_{b,0}(\Omega,q),
\end{align*}
where $\int_{\Omega,q}$ is the frequency-momentum integration for the momentum shell between the UV cutoff ($\Lambda$) and the IR cutoff ($\Lambda e^{-\ell}$). The coupling constants ($g_{\varphi}=ie$, $g_{\phi}=g$) and the operators ($M_{\varphi}=\tau_{0}\sigma_{0}$, $M_{\phi}=\tau_{0}\sigma_{z}$) are introduced with the subscripts. 
The Coulomb interaction (order parameter) self energy can be obtained by evaluating the diagram of Fig.\ref{fig1:c} (Fig.\ref{fig1:d}),
\begin{align*}
\Pi_{b}(\omega,\bm{k})
&=-g_{b}^{2}\int_{\Omega,q}\text{Tr}\left[ M_{b}G_{f,0}(\Omega-\tfrac{\omega}{2},\bm{q}-\tfrac{\bm{k}}{2}) M_{b}G_{f,0}(\Omega+\tfrac{\omega}{2},\bm{q}+\tfrac{\bm{k}}{2}) \right],
\end{align*}
with the boson subscript $b=\varphi, \phi $.
The vertex corrections are from the diagrams of Fig.\ref{fig1:e}, \ref{fig1:f}, \ref{fig1:g}, and \ref{fig1:h},
\begin{align*}
\Gamma_{bb'}=&g_{b'}^{2}\int_{\Omega,q}M_{b'}G_{f,0}(\Omega,\bm{q})M_{b}G_{f,0}(\Omega,\bm{q})M_{b'}G_{b,0}(\Omega,\bm{q}),
\end{align*}
where $\Gamma_{bb'}$ is proportional to $M_{b}$, so it gives the correction to $g_{b}$.
The $\phi^{4}$ coupling constant corrections are from the diagram of  Fig.\ref{fig1:i},
\begin{align*}
\delta\lambda_{1}=\frac{4!\times2}{2!}\left(\frac{\lambda}{4u_{\perp}}\right)^{2}\int_{\Omega,q}G_{b,0}(\Omega,\bm{q})G_{b,0}(\Omega,\bm{q}),
\end{align*}
and  the one of Fig.\ref{fig1:j},
\begin{align*}
\delta\lambda_{2}=&-6g^{4}\int_{\Omega,q}\text{Tr}[G_{f,0}(\Omega,\bm{q})M_{\phi}G_{f,0}(\Omega,\bm{q})M_{\phi}G_{f,0}(\Omega,\bm{q})M_{\phi}G_{f,0}(\Omega,\bm{q})M_{\phi}].
\end{align*}
The numerical factors are from counting all the possible Wick contractions properly.

The one loop corrections may be written in terms of the corrections to the bare action , 
\begin{align*}
\delta S=&\int d^{3}xd\tau\;\left[-\Psi^{\dagger}(\Sigma_{\phi}+\Sigma_{\varphi})\Psi -\frac{1}{2}\phi(\Pi_{\phi})\phi -\frac{1}{2}\varphi(\Pi_{\varphi})\varphi\right]\\
&+\int d^{3}xd\tau\; (\Gamma_{\phi\phi}+\Gamma_{\phi\varphi})g\phi(\Psi^{\dagger}M_{\phi}\Psi)\\
&+\int d^{3}xd\tau\; (\Gamma_{\varphi\varphi}+\Gamma_{\varphi\phi})ie\varphi(\Psi^{\dagger}\Psi)\\
&+\int d^{3}xd\tau\; \frac{1}{4!}(-\delta\lambda_{1}-\delta\lambda_{2})\phi^{4}. 
\end{align*}
Next, we renormalize space-time, $x\rightarrow x e^{\ell}$, $\tau\rightarrow\tau e^{z\ell}$, coupling constants, and wave functions by introducing  $\Psi\rightarrow Z_{\Psi}^{-1/2}\Psi$, $\phi\rightarrow Z_{\phi}^{-1/2}\phi$, $\varphi\rightarrow Z_{\varphi}^{-1/2}\varphi$, $v_{\perp}\rightarrow Z_{v_{\perp}}^{-1}v_{\perp}$, $v_{z}\rightarrow Z_{v_{z}}^{-1}v_{z}$, $u_{\perp}\rightarrow Z_{u_{\perp}}^{-1}u_{\perp}$, $u_{z}\rightarrow Z_{u_{z}}^{-1}u_{z}$, $\gamma\rightarrow Z_{\gamma}^{-1}\gamma$, $g\rightarrow Z_{\alpha_{g}}^{-1/2}g$, $e\rightarrow Z_{\alpha_{e}}^{-1/2}e$, and $\lambda\rightarrow Z_{\lambda}^{-1}\lambda$.
Imposing scale invariance, we may find renormalization of the coupling constants.  

After straightforward calculations, we find the complete RG equations of the six dimensionless parameters, 
\begin{align}\notag
\frac{d R_{\perp}}{d\ell}=& R_{\perp}\left[ \frac{\alpha_{g}}{\pi}\left( \frac{N_{f}}{3}(1- R_{\perp}^{2})+F_{x}( R_{\perp}, R_{z}) \right)-\frac{\alpha_{e}}{\pi}H_{x}( R_{fc}) \right],\\\notag
\frac{d R_{z}}{d\ell}=& R_{z}\left[-\frac{\alpha_{g}}{\pi}\left( \frac{ R_{\perp}^{2}}{3}N_{f}-F_{z}( R_{\perp}, R_{z}) \right)-\frac{\alpha_{e}}{\pi}H_{z}( R_{fc})\right],\\\notag
\frac{d R_{fc}}{d\ell}=& R_{fc}\left[  -\frac{\alpha_{g}}{\pi}\left(F_{z}( R_{\perp}, R_{z})-F_{x}( R_{\perp}, R_{z})\right)\right.\\\notag
&\quad\left.-\frac{\alpha_{e}}{\pi}\left( \frac{N_{f}}{3}( R_{fc}^{2}-1)+(H_{x}( R_{fc})-H_{z}( R_{fc})) \right)\right],\\\notag
\frac{d\alpha_{g}}{d\ell}=&\alpha_{g}\left[ -\frac{\alpha_{g}}{\pi}\left(\frac{2}{3}N_{f}+F_{z}( R_{\perp}, R_{z})\right)+\frac{\alpha_{e}}{\pi}H_{z}( R_{fc}) \right],\\\notag
\frac{d\alpha_{e}}{d\ell}=&\alpha_{e}\left[ -\frac{\alpha_{e}}{\pi}\left(\frac{2}{3}N_{f}+H_{z}( R_{fc})\right)+\frac{\alpha_{g}}{\pi}F_{z}( R_{\perp}, R_{z}) \right],\\
\frac{d\tilde{\lambda}}{d\ell}=&\tilde{\lambda}\left[-\frac{3\tilde{\lambda}}{16\pi^{2}}-\frac{N_{f}}{3\pi}\alpha_{g}(2+ R_{\perp}^{2})\right].\label{eq:beta}
\end{align}
The dimensionless functions ($F_{x}$, $F_{z}$, $H_x$, and $H_z$) are defined as follows, 
\begin{eqnarray}
 F_{x}( a, b)&= &\frac{ a^{2}}{2}\left[ \frac{ b( a^{2}-1)+3( a^{2}- b^{2})}{(1+ b)( a^{2}- b^{2})( a^{2}-1)} -\frac{((2 a^{2}+1) a^{2}-( a^{2}+2) b^{2})}{( a^{2}-1)^{3/2}( a^{2}- b^{2})^{3/2}}\tanh^{-1}\left( \frac{\sqrt{ a^{2}-1}\sqrt{ a^{2}- b^{2}}}{ a^{2}+ b} \right)\right], \label{eq:Fx} \\
 F_{z}( a, b)&= & a^{2}\left[ \frac{ a^{2}- b}{( a^{2}-1)( a^{2}- b^{2})}-\frac{( a^{2}( b^{2}+1)-2 b^{2})}{( a^{2}-1)^{3/2}( a^{2}- b^{2})^{3/2}}\tanh^{-1}\left( \frac{\sqrt{ a^{2}-1}\sqrt{ a^{2}- b^{2}}}{ a^{2}+ b} \right) \right],\label{eq:Fz} \\
H_{x}( c) &=& \frac{c}{2}\left[ \frac{ c}{ c^{2}-1}+\frac{(2c^{2}-1)}{( c^{2}-1)^{3/2}}\tanh^{-1}\left( \frac{\sqrt{ c^{2}-1}}{ c}\right) \right],  \label{eq:Hx}\\
 H_{z}( c)&=&%\frac{1}{\pi}\int_{0}^{\infty}dr\int_{-\infty}^{\infty}dy\;\frac{r(1+r^{2}-y^{2})}{(1+r^{2}+y^{2})^{2}(r^{2}+y^{2}/ c^{2})}\\
 c\left[ \frac{ c}{ c^{2}-1}-\frac{1}{( c^{2}-1)^{3/2}}\tanh^{-1}\left( \frac{\sqrt{ c^{2}-1}}{ c}\right) \right].\label{eq:Hz}
\end{eqnarray} %\end{widetext}
Detailed analysis of the four functions are presented in Appendix \ref{app:loopfunction}.
Note that $F_{x}$ and $F_{z}$ are from the interaction between the order parameter-fermion, and $H_{x}$ and $H_{z}$ are from the Coulomb interaction-fermion loop diagrams.
We also find the following relations,
\begin{align*}
H_{x}(a)=&F_{z}(a,0)-F_{x}(a,0),\\
H_{z}(a)=&F_{z}(a,0).
\end{align*}
And thus, it is enough to investigate $F_{x}$ and $F_{z}$ for the RG analysis. 
We stress that the RG equations are perturbative in the fine structure constants ($\alpha_e, \alpha_g$) but non-perturbative in anisotropic parameters such as velocities in sharp contrast to relativistic quantum field theories where anisotropy is forbidden by the Lorentz symmetry. 
Thus, we may access quantum criticalities with strong anisotropy.

In the RG equations, the first thing we emphasize is that all the coupling constants are marginally irrelevant giving $\alpha_{e}(l), \alpha_{g}(l), \lambda(l) \propto l^{-1}$ in the long wavelength limit, $l \rightarrow \infty$ demonstrating weakly coupled fixed points if they are stable. 
The remaining RG equations of $R_z$, $R_{\perp}$, and $R_{fc}$ can be more manifestly analyzed by introducing the anisotropy parameters,  $\gamma_{f}\equiv v_{z}/v_{\perp} = R_{fc} \gamma_c$, $\gamma_{b}\equiv u_{z}/u_{\perp} = \frac{R_z}{R_{\perp}} R_{fc} \gamma_c$ and $\gamma_{c}$. 
The flow equations of the anisotropy constants are as follows:
\begin{align}
\frac{d\gamma_{f}}{d\ell}\equiv&\left\{-\frac{\alpha_{g}}{\pi}\left[ F_{z}( R_{\perp}, R_{z})-F_{x}( R_{\perp}, R_{z}) \right]\right.\notag\\
&\left.\quad+\frac{\alpha_{e}}{\pi}\left[ H_{z}( R_{fc})-H_{x}( R_{fc}) \right]\right\}\gamma_{f},\label{eq:gammaf}\\
\frac{d\gamma_{b}}{d\ell}\equiv&-\frac{N_{f}}{3\pi}\alpha_{g}\gamma_{b},\label{eq:gammab}\\
\frac{d\gamma_{c}}{d\ell}=&-\frac{N_{f}}{3\pi}\alpha_{e}(1- R_{fc}^{2})\gamma_{c}.\label{eq:gammac}
\end{align}
Below, we present our RG analysis results in turn :  1) the case with fermion and the order parameter, 2) the case with fermions, order parameter, and the Coulomb interaction, and 3) non-Fermi liquid phase with the order parameter.
To be self-contained, we present the case with fermions and the Coulomb interaction in appendix \ref{app:coulombrg}.

\begin{figure}
\centering
\subfigure[]{
\includegraphics[width=0.472\linewidth]{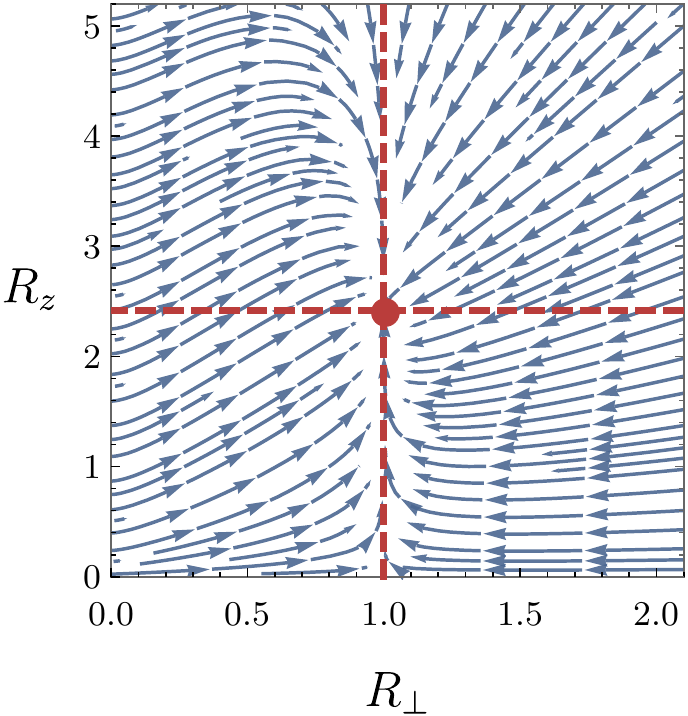}\label{fig:xzn1}
}\hfill
\subfigure[]{
\includegraphics[width=0.472\linewidth]{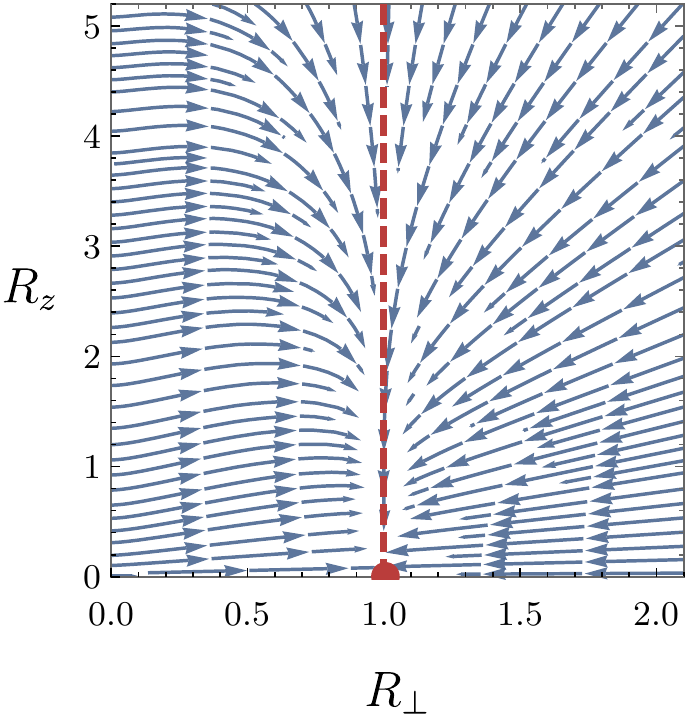}\label{fig:xzn2}
}
\caption{RG flows and fixed points for $ R_{\perp}$ and $ R_{z}$ without Coulomb interaction in terms of $N_{f}$. (a) RG flow and fixed point of $ R_{\perp}$ and $ R_{z}$ for $N_{f}=1$. The fixed point value (red dot) is $( R_{\perp}^{*}, R_{z}^{*})=(1,1+\sqrt{2})$. (b) RG flow and fixed point of $ R_{\perp}$ and $ R_{z}$ for $N_{f}=2$. The fixed point value (red dot) is $( R_{\perp}^{*}, R_{z}^{*})=(1,0)$. For $N_{f}\geq2$, the fixed point value and RG flow are similar to $N_{f}=2$ case. Note that these RG flows are drawn with the value $\alpha_{g}=1$.}\label{fig:rg_flow_woc}
\end{figure}

\subsection{Fermion and order parameter} \label{without}
Let us first consider the case without the Coulomb interaction. This case may be naturally realized in  a phase transition between the Dirac and Weyl superconductors \cite{PhysRevLett.113.046401} and a transition between Dirac and Weyl semi-metals with large Coulomb screening. 
Starting with $\alpha_e=0$, it is enough to take into account the RG equations of $ R_{\perp}$, $ R_{z}$, $\alpha_{g}$, and $\tilde{\lambda}$,
\begin{align}\notag
\frac{d R_{\perp}}{d\ell}=& R_{\perp}\left[ \frac{\alpha_{g}}{\pi}\left( \frac{N_{f}}{3}(1- R_{\perp}^{2})+F_{x}( R_{\perp}, R_{z}) \right)  \right],\\\notag
\frac{d R_{z}}{d\ell}=& R_{z}\left[-\frac{\alpha_{g}}{\pi}\left( \frac{ R_{\perp}^{2}}{3}N_{f}-F_{z}( R_{\perp}, R_{z}) \right) \right],\\\notag
\frac{d\alpha_{g}}{d\ell}=&\alpha_{g}\left[ -\frac{\alpha_{g}}{\pi}\left(\frac{2}{3}N_{f}+F_{z}( R_{\perp}, R_{z})\right)  \right],\\\notag
\frac{d\tilde{\lambda}}{d\ell}=&\tilde{\lambda}\left[-\frac{3\tilde{\lambda}}{16\pi^{2}}-\frac{N_{f}}{3\pi}\alpha_{g}(2+ R_{\perp}^{2})\right].
\end{align}
By using Fig.\ref{fig:fxplot} and Eq.\ref{eq:cond1} in Appendix \ref{app:loopfunction}, we find that $R_{\perp}=1$ is a necessary condition to be a fixed point. 
Thus, along the perpendicular directions, the fermion and boson velocities become same. 
At $ R_{\perp}=1$, the remaining equations become
\begin{align*}
\frac{d R_{z}}{d\ell}=&\frac{\alpha_{g}}{3\pi}\left( \frac{2(1+2 R_{z})}{(1+ R_{z})^{2}} - N_{f}\right) R_{z},\\
\frac{d\alpha_{g}}{d\ell}=& -\frac{2\alpha_{g}^{2}}{3\pi}\left(N_{f}+\frac{1+2 R_{z}}{(1+ R_{z})^{2}}\right),\\
\frac{d\tilde{\lambda}}{d\ell}=&-\frac{3\tilde{\lambda}^{2}}{16\pi^{2}}-\frac{N_{f}}{\pi}\tilde{\lambda}\alpha_{g}.
\end{align*}
Since $\alpha_{g}$ is positive semi-definite, the fixed point values of the coupling constants are obviously $(\alpha_{g}^{*},\tilde{\lambda}^{*})=(0,0)$. 
Remark that the fixed point value of $R_z$ depends on $N_{f}$. For $N_{f}=1$, $ R_{z}^{*}=1+\sqrt{2}$ while $ R_{z}^{*}=0$ for $N_{f}\geq2$ (see Fig.\ref{fig:fzplot}). The RG flows with $ R_{\perp}$ and $ R_{z}$ are illustrated in Fig.\ref{fig:rg_flow_woc}.

Let us further analyze the RG equations. 
The fine structure constant is marginally irrelevant, $\frac{d}{d l} \alpha_g \propto - \alpha_g^2$, which gives $\alpha_g(l) \propto l^{-1}$. Then, the boson anisotropy RG equation makes $\gamma_b \rightarrow 0$. Moreover, the condition $F_z - F_x \ge 0$ (Appendix \ref{app:loopfunction}) makes the fermion anisotropy vanish, $\gamma_f \rightarrow 0$ in the long wave length limit. 
Therefore, the fermion and boson excitations become infinitely anisotropic. 

The fixed points for $N_f=1$ and $N_f \ge 2$ demonstrate the presence of the two different types of quantum criticalities with infinite anisotropy. 
For $N_f=1$, the non-unity value of $R_z^* = 1+\sqrt{2}$ indicates that the boson and fermion excitations move with the velocity ration along the z direction (nodal point splitting direction). 
On the other hand, for $N_f \ge 2$, the fermion excitations move qualitatively faster than the boson excitations along the $z$ direction. Note that both cases show infinite anisotropy in both fermion and boson excitations at quantum critical points.
 
\begin{figure}
\centering
\subfigure[]{
\includegraphics[]{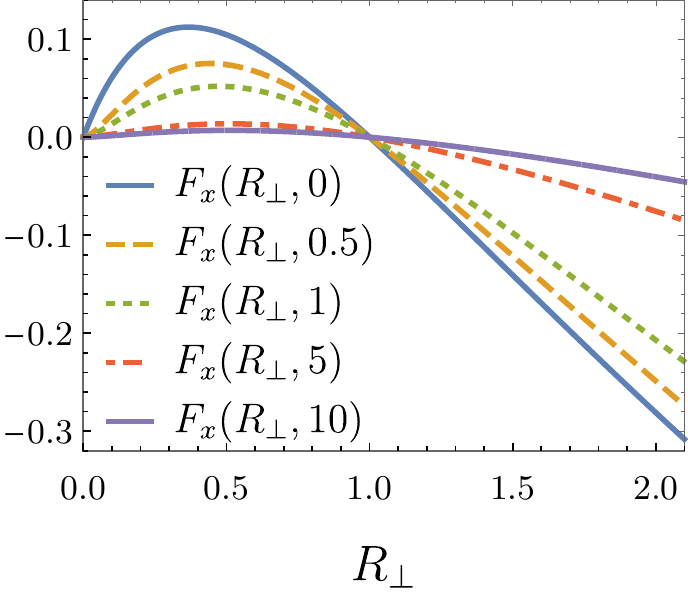}\label{fig:fxplot}
}\hfill
\subfigure[]{
\includegraphics[]{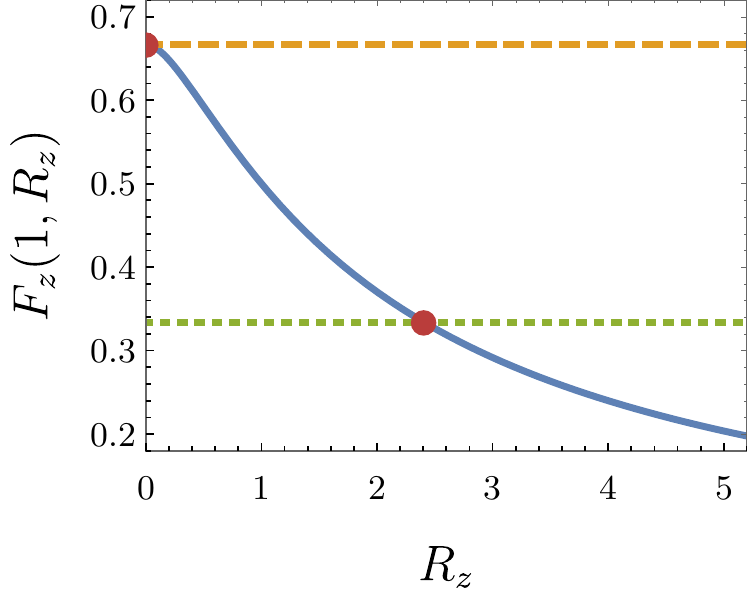}\label{fig:fzplot}
}
\caption{  Behaviors of $F_{x}$ and $F_{z}$. (a) The behavior of $F_{x}$. $F_{x}( R_{\perp}, R_{z})>0$ for $ R_{\perp}<1$, and $F_{x}( R_{\perp}, R_{z})<0$ for $ R_{\perp}>1$ regardless of $ R_{z}$. (b) The behavior of $F_{z}( R_{\perp}, R_{z})$ for $ R_{\perp}=1$. The blue solid line is for $F_{z}(1, R_{z})$, and the dashed (dotted) orange (green) line is $1/3$ (2/3). $F_{z}(1, R_{z})-1/3>0$ for $ R_{z}<1+\sqrt{2}$, but $F_{z}(1, R_{z})-1/3<0$ for $ R_{z}>1+\sqrt{2}$. However, $F_{z}(1, R_{z})-2/3\leq0$ regardless of $ R_{z}$.
}\label{fig:ffunction}
\end{figure}

\subsection{Fermion, order parameter, and Coulomb interaction} \label{with}

\begin{figure}
\centering
\includegraphics[]{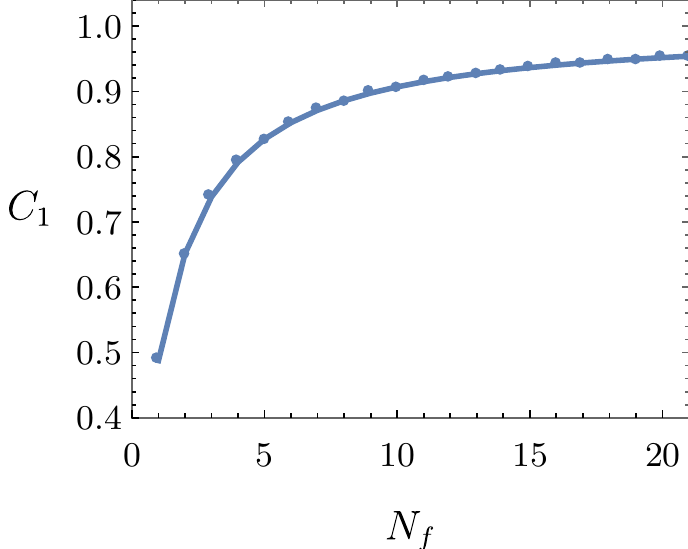}
\caption{ 
$C_{1}$ as function of $N_{f}$. we can see that $C_{1}$ strictly increases to 1 when $N_{f}\rightarrow\infty$.}\label{fig:c1plot}
\end{figure}

Let us consider the case with the long-range Coulomb interaction which is naturally realized in the Dirac semi-metal to Weyl semi-metal transition without large Coulomb screening. 
At the lattice scale, the fine structure constant ($\alpha_e$) is not negligible and we should keep it from the beginning. From the full RG equations, we obtain a stable fixed point, $( R_{\perp}^{*}, R_{z}^{*}, R_{fc}^{*},\alpha_{g}^{*},\alpha_{e}^{*},\tilde{\lambda})=(C_{1}(N_f),0,C_{1}(N_f),0,0,0)$. 
The numerical value of $C_1(N_f)$ is determined in Fig. \ref{fig:c1plot}. 
In contrast to the case without the Coulomb interaction, the numerical value of  $ R_{\perp}^{*}$ is smaller than the unity. We note that similar suppression of the velocity ration is also reported in the quantum phase transitions between semi-metals and insulators with the long-range Coulomb interaction \cite{2018arXiv180205727H}. 
Interestingly, we find that $ R_{z}$, in contrary to the case without the Coulomb interaction,  vanishes for all $N_{f}$ near fixed point. It is because that the Coulomb interaction makes the fermion faster while the bosons are not directly coupled to the Coulomb interaction. Thus, the ratio $R_z^*$ is more suppressed.  

We consider the ratio, $\mathcal{R}_{\alpha}\equiv\alpha_{e}/\alpha_{g}$. Without the Coulomb interaction, obviously $\mathcal{R}_{\alpha}(\alpha_{e}=0)=0$. But allowing the Coulomb interaction at the microscopic level, the situation is changed significantly in the long wave-length limit. From the RG flow equations of $\alpha_{g}$ and $\alpha_{e}$, we find that (Appendix \ref{app:couplingratio})
\begin{align*}
\mathcal{R}_{\alpha}\rightarrow \frac{N_{f}+3 F_{z}( R_{\perp}^{*}, R_{z}^{*})}{N_{f}+3H_{z}( R_{fc}^{*})}.
\end{align*}
From Eq.\ref{eq:Fz} and Eq.\ref{eq:Hz}, we know that $F_{z}( R_{\perp},0)$ and $H_{z}( R_{fc})$ have the same form when $ R_{z}=0$ (Appendix \ref{app:loopfunction}). Therefore, near the fixed point $( R_{\perp}^{*}, R_{z}^{*}, R_{fc}^{*})=(C_{1}(N_{f}),0,C_{1}(N_{f}))$, the coupling ratio $\mathcal{R}_{\alpha}$ becomes 1 independent of $N_{f}$.

The anisotropic parameters show characteristic behaviors under the long-range Coulomb interaction.
From Eq.\ref{eq:gammab}, $\gamma_{b}$ vanishes, so the order parameter again becomes infinitely anisotropic.
Using $\mathcal{R}_{\alpha}\rightarrow1$ and fixed-point values, we know that $\gamma_{f}$ and $\gamma_{c}$ also vanish, so the fermion and the Coulomb interaction become anisotropic. 
Thus, all the excitations and the long-range Coulomb interaction become anisotropic. 
 
Note that when $N_{f}\rightarrow\infty$, the factor $(1- R_{fc}^{2})$ approaches to 0, so it seems that $\frac{d\gamma_{c}}{d\ell}=0$ with a non-zero $\gamma_{c}$. However, in that limit, since $N_{f}$ and $(1- R_{fc}^{2})$ are balanced, $\gamma_{c}$ vanishes even though $ R_{fc}\rightarrow1$ when $N_{f}\rightarrow\infty$. In other words,  the Coulomb interaction is still anisotropic. This may be seen from $ R_{fc}=\gamma_{f}/\gamma_{c}$ and $\gamma_{f}\rightarrow0$ independent of $N_{f}$.
This is similar to the previously studied case of the quantum phase transition between the non-Fermi liquid and Weyl semi-metal  \cite{PhysRevX.4.041027} in a sense that all excitations become anisotropic under the presence of the Coulomb interaction.

\subsection{Non-Fermi liquid phase} \label{NFL}
Our RG analysis may be further applied to the case with non-Fermi liquid by assuming the presence of a fixed point with a non-zero value $\alpha_e^*$ with $\left.\frac{d\alpha_{e}}{d\ell}\right|_{\alpha_{e}=\alpha_{e}^{*}}=0$. 
If the fixed point is controlled by a small parameter, say $1/N_{f}$, then we may extend our weak coupling analysis to strongly coupled fixed points. Under the assumption of the presence of $\alpha_e^* \neq0$,  we find a stable fixed point value, $( R_{\perp}^{*}, R_{z}^{*}, R_{fc}^{*},\alpha_{g}^{*},\alpha_{e}^{*},\tilde{\lambda}^{*})=(0,0,1,\alpha_{e}^{*}/N_{f},\alpha_{e}^{*},0)$. 
Note that $R_{fc}^{*}=1$ indicates that the two anisotropy values ($\gamma_{f}$ and $\gamma_{c}$) are the same whose value, $\gamma_{0}$, is determined by the values at the microscopic scale ($\gamma_{f}(l=0)$ and $\gamma_{c}(l=0)$).  

It is interesting that $R_{fc}^{*}=1$ in the non-Fermi liquid phase, and we believe this is intrinsically tied to the fact that our system becomes strongly coupled.  Note that  $\gamma_{b}\rightarrow0$, indicating the order parameter is infinitely anisotropic near the fixed point. We believe our results may be naturally connected to strongly coupled fixed points with infinite anisotropy.

\begin{table}
\begin{tabular}{c|@{}>{$}c<{$}@{} @{}>{$}c<{$}@{}  @{}>{$}c<{$}@{}|@{}>{$}c<{$}@{}  @{}>{$}c<{$}@{}  @{}>{$}c<{$}@{}  @{}>{$}c<{$}@{}  @{}>{$}c<{$}@{}  @{}>{$}c<{$}@{} }
\hline\hline
 Cases &~~\gamma_{f}~~&~~\gamma_{b}~~&~~\gamma_{c}~~&~~~R_{\perp}~~~&~~~R_{z}~~~&~~~R_{fc}~~~&~~~~\alpha_{g}~~~~&~~~~\alpha_{e}~~~~&~~~~\tilde{\lambda}~~~~\\\hline\hline
 %only Coulomb&\gamma_{0}& &\gamma_{0}& & &1& &0& \\\cline{1-1}
 w/o Coulomb (\ref{without}) \begin{tabular}{|@{}>{$}c<{$}@{}}~N_{f}=1~\\ ~N_{f}\geq2~\\\end{tabular}&\begin{tabular}{@{}>{$}c<{$}@{}}0\\0\end{tabular}&\begin{tabular}{@{}>{$}c<{$}@{}}0\\0\end{tabular}& &\begin{tabular}{@{}>{$}c<{$}@{}}1\\1\end{tabular}&\begin{tabular}{@{}>{$}c<{$}@{}}1+\sqrt{2}\\0\end{tabular}& &\begin{tabular}{@{}>{$}c<{$}@{}}0\\0\end{tabular}& &\begin{tabular}{@{}>{$}c<{$}@{}}0\\0\end{tabular}\\\cline{1-1}
w/ Coulomb (\ref{with}) &0&0&0&C_{1}(N_{f})&0&C_{1}(N_{f})&0&0&0\\\cline{1-1}
NFL (\ref{NFL}) &\gamma_{0}&0&\gamma_{0}&0&0&1&\alpha_{e}^{*}/N_{f}&\alpha_{e}^{*}&0\\\hline
\end{tabular}
\caption{Summary of the dimensionless parameters of the subsections (\ref{without}, \ref{with}, \ref{NFL}). The blank cells are for ill-defined parameters, and the explicit value of $C_{1}(N_{f})$ is presented at Fig.\ref{fig:c1plot}.
}
\end{table}

\section{Discussion and Conclusion}
We discuss implications of our findings. First, the topological phase transitions associated with Fermi-point splitting in three spatial dimensions are characteristically {\it correlated} in a sense that the order parameter dynamics becomes infinitely anisotropic. Surprisingly, the order parameter does not receive any anomalous dimensions in spite of the presence of infinite anisotropy. This is in sharp contrast to the previously suggested quantum criticalities with infinite anisotropy where a large anomalous dimension and the infinite anisotropy appear simultaneously \cite{PhysRevB.78.064512,PhysRevX.4.041027}. Our perturbative RG calculations manifestly show that the infinite anisotropy and the presence of a non-zero anomalous dimension are not tied. 
Second, characteristic behaviors of coupling constants at low energy may be measured in various experiments. For example, the optical conductivity $\sigma_{ij}$ is proportional to $v_{i}v_{j}$ and $\gamma_{f}\equiv v_{z}/v_{\perp}$ vanishes near the fixed point, thus the ratio $\sigma_{xx}/\sigma_{zz}$, which is proportional to $\gamma_{f}^{-2}$, may be tested in experiments at low temperature.  
Third, our RG analysis indicates that the fixed points we find in this paper may be naturally connected to the previously suggested strongly coupled fixed points. By assuming the presence of a non-zero coupling constant of the fine structure constant ($\alpha_e^*$), we find different features of a strongly coupled fixed point even though it has infinite anisotropy. Thus, it is natural to interpret our weakly coupled fixed points as a ``Gaussian'' fixed point with infinite anisotropy. 
 
In conclusion, we study quantum phase transitions (QPTs) associated with splitting nodal Fermi points, motivated by topological phase transitions between Dirac and Weyl semi-metals. Striking correlation effects at quantum critical points such as { infinite anisotropy} of physical quantities are obtained by using the RG analysis. For example, we show the presence of weakly coupled quantum criticalities with infinite anisotropy.  Our results demonstrate that correlation effects should be considered from the beginning in topological phase transitions between Dirac and Weyl semi-metals.  

\begin{acknowledgments}
We thank S. S. Lee for invaluable discussion. This work was supported by the POSCO Science Fellowship of POSCO TJ Park Foundation and NRF of Korea under Grant No. 2017R1C1B2009176. GYC was in part supported from the Korea Institute for Advanced Study (KIAS) grant funded by the Korea government (MSIP).
\end{acknowledgments}

\appendix

\section{Outline of calculations}

From Eq.\ref{eq:action}, we have the partition function,
\begin{align}
Z=\int D\Psi D\phi D\varphi e^{-\mathcal{S}}.
\end{align}
Let $\mathcal{S}_{0}$ be the action for the quadratic terms, and $\mathcal{S}_{1}$ be the action for the interactions, so $\mathcal{S}=\mathcal{S}_{0}+\mathcal{S}_{1}$. Dividing each field into the fast modes ($\Psi_{>}$, $\phi_{>}$, and $\varphi_{>}$), and slow modes ($\Psi_{<}$, $\phi_{<}$, and $\varphi_{<}$), the partition function can be written as
\begin{align*}
Z=&\int D\Psi_{<}D\phi_{<}D\varphi_{<} e^{-\mathcal{S}[\Psi_{<},\phi_{<},\varphi_{<}]}\\
&\times\int D\Psi_{>}D\phi_{>}D\varphi_{>} e^{-\mathcal{S}[\Psi_{>},\phi_{>},\varphi_{>}] -\mathcal{S}_{\text{int}}[\Psi_{<},\Psi_{>},\phi_{<},\phi_{>},\varphi_{<},\varphi_{>}]}\\
=&Z_{>}\int D\Psi_{<}D\phi_{<}D\varphi_{<} e^{-\mathcal{S}[\Psi_{<},\phi_{<},\varphi_{<}]} \braket{e^{-\mathcal{S}_{\text{int}}}}_{>},
\end{align*}
where $Z_{>}=\int D\Psi_{>}D\phi_{>}D\varphi_{>} e^{-\mathcal{S}[\Psi_{>},\phi_{>},\varphi_{>}]}$ is the normalization factor about the fast mode integration, $\mathcal{S}_{\text{int}}$ is the action for the interactions between the fast and slow modes,
\begin{align*}
&\mathcal{S}_{\text{int}}[\Psi_{<},\Psi_{>},\phi_{<},\phi_{>},\varphi_{<},\varphi_{>}]\\
&\equiv\mathcal{S}_{1}[\Psi_{<}+\Psi_{>},\phi_{<}+\phi_{>},\varphi_{<}+\varphi_{>}]\\
&\quad-\mathcal{S}_{1}[\Psi_{<},\phi_{<},\varphi_{<}]-\mathcal{S}_{1}[\Psi_{>},\phi_{>},\varphi_{>}],
\end{align*}
and $\braket{e^{-\mathcal{S}_{\text{int}}}}_{>}$ means the average of $e^{-\mathcal{S}_{\text{int}}}$ in terms of the fast modes integration. $\braket{e^{-\mathcal{S}_{\text{int}}}}_{>}$ can be obtained by
\begin{align*}
\braket{e^{-\mathcal{S}_{\text{int}}}}_{>}\equiv&\frac{1}{Z_{>}}\int D\Psi_{>}D\phi_{>}D\varphi_{>} e^{-\mathcal{S}[\Psi_{>},\phi_{>},\varphi_{>}]}e^{ -\mathcal{S}_{\text{int}}}\\
=&1-\braket{S_{\text{int}}}_{>}+\frac{1}{2!}\braket{S_{\text{int}}^{2}}_{>}-\frac{1}{3!}\braket{S_{\text{int}}^{3}}_{>}+\frac{1}{4!}\braket{\mathcal{S}_{\text{int}}^{4}}_{>}+\cdots\\
\approx&e^{-\left[ \braket{S_{\text{int}}}_{>}-\frac{1}{2!}\braket{S_{\text{int}}^{2}}_{>}+\frac{1}{3!}\braket{S_{\text{int}}^{3}}_{>}-\frac{1}{4!}\braket{\mathcal{S}_{\text{int}}^{4}}_{>}+\cdots \right]}\\
=&e^{-\delta\mathcal{S}},
\end{align*}
where $\braket{\mathcal{S}_{\text{int}}^{i}}_{>}$ is the $i$-th cumulant expansion in terms of the fast modes integration, $\delta\mathcal{S}$ is the leading order correction for the action.\\\indent
The self energies of the fermion, order parameter, and Coulomb interaction, and the quartic vertex correction by the order parameter loop come from $\braket{\mathcal{S}_{\text{int}}^{2}}$. The order parameter- fermion vertex correction and the quartic vertex correction by the fermion loop come from $\braket{\mathcal{S}_{\text{int}}^{3}}$ and $\braket{\mathcal{S}_{\text{int}}^{4}}$, respectively.
% The each corrections are as follows.\\\indent
%The fermion self energy is given by Fig.\ref{fig1:a}, \ref{fig1:b} at the one loop order,
%\begin{align*}
%&\Sigma_{b}(\omega,\bm{k})\\
%&=g_{b}^{2}\int_{\Omega,q}M_{b}G_{f,0}(\omega+\Omega,\bm{k}+\bm{q})M_{b}G_{b,0}(\Omega,q),
%\end{align*}
%where $\int_{\Omega,q}$ is the frequency-momentum integration for the momentum shell between $\Lambda e^{-\ell}$ and $\Lambda$, $g_{\varphi}=ie$, $g_{\phi}=g$, $M_{\varphi}=\tau_{0}\sigma_{0}$ is for instantaneous Coulomb interaction, $M_{\phi}=\tau_{0}\sigma_{z}$ are the coupling matrices for the symmetry breaking order parameters interaction which is splitting the Dirac point into the Weyl points along $k_{z}$ axis. 
%$G_{f,0}$, $G_{\phi,0}$, and $G_{\varphi,0}$ are the bare propagators of the fermion, order parameter and Coulomb interaction which are defined in main text.
%
% by
%\begin{align}
%G_{f,0}(\omega,\bm{k})=&\frac{1}{-i\omega+\mathcal{H}_{0}(\bm{k})}\notag\\=&\frac{i\omega+\tau_{z}(v_{\perp}(\sigma_{x}k_{x}+\sigma_{y}k_{y})+v_{z}\sigma_{z}k_{z})}{\omega^{2}+v_{\perp}^{2}k_{\perp}^{2}+v_{z}^{2}k_{z}^{2}},\\
%G_{\varphi,0}(\omega,\bm{k})=&\frac{1}{k_{\perp}^{2}+\gamma_{c}^{2}k_{z}^{2}},\\
%G_{\phi,0}(\omega,\bm{k})=&\frac{1}{\omega^{2}/u_{\perp}^{2}+k_{\perp}^{2}+(u_{z}/u_{\perp})^{2}k_{z}^{2}}.%,
%\end{align}

\section{Independence of choice of cutoff axis}\label{sm:rgscheme}
In this section, we discuss the RG scheme independence. In the main text, we used the shell integral for the frequency, i.e., $\int_{\Lambda/b}^{\Lambda}d\Omega\int_{-\infty}^{\infty}d^{3}p$. Here, we apply the shell integral at momenta, $q_{x}$ and $q_{z}$.\\
For shell integration about $q_{x}$ and $q_{z}$, they give us the same structure, but different loop function. For $q_{x}$ integration, i.e., $\int_{-\infty}^{\infty}d\Omega\int_{\Lambda/b}^{\Lambda}dq_{x}\int_{-\infty}^{\infty}dq_{y}dq_{z}$, 
each loop function replace by
\begin{align*}
F_{x}(a,b)&\rightarrow \xi_{1}(a,b),\\
F_{z}(a,b)&\rightarrow \xi_{2}(a,b),\\
H_{x}(c)&\rightarrow \Xi_{x}(c),\\
H_{z}(c)&\rightarrow \Xi_{z}(c),
\end{align*}
where
\begin{align*}
\xi_{x}(a,b)\equiv&\frac{1}{\pi^{2}}\int_{-\infty}^{\infty}\frac{1-w^{2}}{(1+w^{2}+y^{2}+z^{2})^{2}(1+w^{2}/a^{2}+y^{2}+(b^{2}/a^{2})z^{2})}dwdydz\\
=&\frac{a^{2}}{2}\left[ \frac{b(a^{2}-1)+3(a^{2}-b^{2})}{(1+b)(a^{2}-b^{2})(a^{2}-1)} -\frac{((2a^{2}+1)a^{2}-(a^{2}+2)b^{2})}{(a^{2}-1)^{3/2}(a^{2}-b^{2})^{3/2}}\tanh^{-1}\left( \frac{\sqrt{a^{2}-1}\sqrt{a^{2}-b^{2}}}{a^{2}+b} \right)\right],\\
\xi_{z}(a,b)\equiv&\frac{1}{\pi^{2}}\int_{-\infty}^{\infty}\frac{1+y^{2}}{(1+w^{2}+y^{2}+z^{2})^{2}(1+w^{2}/a^{2}+y^{2}+(b^{2}/a^{2})z^{2})}dwdydz\\
=&a^{2}\left[ \frac{a^{2}-b}{(a^{2}-1)(a^{2}-b^{2})}-\frac{(a^{2}(b^{2}+1)-2b^{2})}{(a^{2}-1)^{3/2}(a^{2}-b^{2})^{3/2}}\tanh^{-1}\left( \frac{\sqrt{a^{2}-1}\sqrt{a^{2}-b^{2}}}{a^{2}+b} \right) \right],\\
\Xi_{1}(c)\equiv&\frac{1}{2\pi^{2}}\int_{-\infty}^{\infty}\frac{w^{2}-1+y^{2}+z^{2}}{(1+w^{2}+y^{2}+z^{2})^{2}(1+y^{2}+z^{2}/c^{2})}dwdydz\\
=&\frac{c}{2}\left[ \frac{c}{1-c^{2}}+\frac{(2c^{2}-1)}{(c^{2}-1)^{3/2}}\tanh^{-1}\left( \frac{\sqrt{c^{2}-1}}{c} \right) \right],\\
\Xi_{2}(c)\equiv&\frac{1}{2\pi^{2}}\int_{-\infty}^{\infty}\frac{w^{2}+1+y^{2}-z^{2}}{(1+w^{2}+y^{2}+z^{2})^{2}(1+y^{2}+z^{2}/c^{2})}dwdydz\\
=&c\left[ \frac{c}{c^{2}-1}-\frac{1}{(c^{2}-1)^{3/2}}\tanh^{-1}\left( \frac{\sqrt{c^{2}-1}}{c}\right) \right].
\end{align*}
From the results of integrations, clearly we have
\begin{align*}
 \xi_{1}(a,b)&\equiv F_{x}(a,b),\\
 \xi_{2}(a,b)&\equiv F_{z}(a,b),\\
 \Xi_{x}(c)&\equiv H_{x}(c),\\
 \Xi_{z}(c)&\equiv H_{z}(c).
\end{align*}
So, we obtain the same result as the main text in this RG scheme.\\
For $q_{z}$ integration, i.e., $\int_{-\infty}^{\infty}d\Omega\int_{-\infty}^{\infty}dq_{y}dq_{z}\int_{\Lambda/b}^{\Lambda}dq_{z}$, each loop function replace by
\begin{align*}
F_{x}(a,b)&\rightarrow \eta_{1}(a,b),\\
F_{z}(a,b)&\rightarrow \eta_{2}(a,b),\\
H_{x}(c)&\rightarrow \kappa_{x}(c),\\
H_{z}(c)&\rightarrow \kappa_{z}(c),
\end{align*}
where
\begin{align*}
\eta_{x}(a,b)\equiv&\frac{1}{\pi}\int_{0}^{\infty}dr\int_{-\infty}^{\infty}dz\frac{r(r^{2}-2z^{2})}{(1+r^{2}+z^{2})^{2}(1+r^{2}(a^{2}/b^{2})+z^{2}/b^{2})}\\
=&\frac{a^{2}}{2}\left[ \frac{b(a^{2}-1)+3(a^{2}-b^{2})}{(1+b)(a^{2}-b^{2})(a^{2}-1)} -\frac{((2a^{2}+1)a^{2}-(a^{2}+2)b^{2})}{(a^{2}-1)^{3/2}(a^{2}-b^{2})^{3/2}}\tanh^{-1}\left( \frac{\sqrt{a^{2}-1}\sqrt{a^{2}-b^{2}}}{a^{2}+b} \right)\right],\\
\eta_{z}(a,b)\equiv&\frac{1}{\pi}\int_{0}^{\infty}dr\int_{-\infty}^{\infty}dz\frac{2r^{3}}{(1+r^{2}+z^{2})^{2}(1+r^{2}(a^{2}/b^{2})+z^{2}/b^{2})}\\
=&a^{2}\left[ \frac{a^{2}-b}{(a^{2}-1)(a^{2}-b^{2})}-\frac{(a^{2}(b^{2}+1)-2b^{2})}{(a^{2}-1)^{3/2}(a^{2}-b^{2})^{3/2}}\tanh^{-1}\left( \frac{\sqrt{a^{2}-1}\sqrt{a^{2}-b^{2}}}{a^{2}+b} \right) \right],\\
\kappa_{1}(c)\equiv&\frac{a^{2}/b^{2}}{\pi}\int_{0}^{\infty}dr\int_{-\infty}^{\infty}dz\frac{r(1+z^{2})}{(1+r^{2}+z^{2})^{2}(r^{2}+c^{-2})}\\
=&\frac{c}{2}\left[ \frac{c}{1-c^{2}}+\frac{(2c^{2}-1)}{(c^{2}-1)^{3/2}}\tanh^{-1}\left( \frac{\sqrt{c^{2}-1}}{c} \right) \right],\\
\kappa_{2}(c)\equiv&\frac{a^{2}/b^{2}}{\pi}\int_{0}^{\infty}dr\int_{-\infty}^{\infty}dz\frac{r(-1+r^{2}+z^{2})}{(1+r^{2}+z^{2})^{2}(r^{2}+c^{-2})}\\
=&c\left[ \frac{c}{c^{2}-1}-\frac{1}{(c^{2}-1)^{3/2}}\tanh^{-1}\left( \frac{\sqrt{c^{2}-1}}{c}\right) \right].
\end{align*}
However we can know that 
\begin{align*}
 \eta_{x}(a,b)&\equiv F_{x}(a,b),\\
 \eta_{z}(a,b)&\equiv F_{z}(a,b),\\
 \kappa_{x}(c)&\equiv H_{x}(c),\\
 \kappa_{z}(c)&\equiv H_{z}(c).
\end{align*}
So, we obtain the same result as the main text in this RG scheme.\\
From the above results, we conclude that our results are independent of the choice of the shell integral for the frequency $\omega$, and the momentum along $q_{x}$ and $q_{z}$.

\section{Proof of asymptotic behavior of $\alpha_{e}/\alpha_{g}$}\label{app:couplingratio}
In this section, we will prove that $\alpha_{e}/\alpha_{g}$ is constant near the fixed point.\\
Let us consider the flow equations of the coupling constants $\alpha_{g}$ and $\alpha_{e}$. Near the fixed point, the coefficients of the flow equations become constants. Then, they has the forms
\begin{align*}
\frac{1}{\alpha_{g}}\frac{d\alpha_{g}}{d\ell}=&-A\alpha_{g}+B\alpha_{e},\\
\frac{1}{\alpha_{e}}\frac{d\alpha_{e}}{d\ell}=&C\alpha_{g}-D\alpha_{e},
\end{align*}
where $A$, $B$, $C$, and $D$ are positive. 
From the RG flow equations of $\alpha_{g}$ and $\alpha_{e}$, the flow equation of the ratio between $\alpha_{g}$ and $\alpha_{e}$ is
\begin{align*}
\frac{d\ln(\alpha_{e}/\alpha_{g})}{d\ell}=&-(B+D)\alpha_{e}+(A+C)\alpha_{g}\\
=&\left(-\frac{B+D}{A+C}(\alpha_{e}/\alpha_{g})+1\right)(A+C)\alpha_{e}.
\end{align*}
Solving this, we have
\begin{align*}
\alpha_{e}/\alpha_{g}=&\left( \frac{B+D}{A+C}+Fe^{-(A+C)\int_{1}^{\ell}\alpha_{e}(x)dx} \right)^{-1}
\end{align*}
where $F$ is a positive constant. Since because $d\alpha_{e}/d\ell\sim-\alpha_{e}^{2}$, let us assume that $\alpha_{e}\sim c_{e}/\ell$ when $\ell\rightarrow\infty$ where $c_{e}>0$. The, $e^{-(A+C)\int_{1}^{\ell}\alpha_{e}(x)dx}\sim e^{-(A+C)c_{e}\ln \ell}\rightarrow0$ as $\ell\rightarrow\infty$. Therefore, when $\ell\rightarrow\infty$, $\alpha_{e}/\alpha_{g}\rightarrow\frac{A+C}{B+D}$. \\
{\it\bfseries Another way of proof}\\
Let us assume the asymptotic behaviors of $\alpha_{g}$ and $\alpha_{e}$ as follows: $\alpha_{g}\simeq c_{g}/\ell$ and $\alpha_{e}\simeq c_{e}/\ell$ for large $\ell$ where $c_{e,g}>0$ because $d\alpha_{g,e}/d\ell\sim-\alpha_{g,e}^{2}$. Then, above equations become
\begin{align*}
-\frac{c_{g}}{\ell^{2}}=&\frac{1}{\ell^{2}}\left(-Ac_{g}^{2}+Bc_{g}c_{e}\right),\\
-\frac{c_{e}}{\ell^{2}}=&\frac{1}{\ell^{2}}\left(C c_{g}c_{e}-Dc_{e}^{2}\right).
\end{align*}
Solving these, we obtain
\begin{align*}
c_{g}=\frac{B+D}{AD-BC},\quad c_{e}=\frac{A+C}{AD-BC}.
\end{align*}
Using these, we can obtain the asymptotic behaviors of $\alpha_{e}/\alpha_{g}$,
\begin{align*}
\alpha_{e}/\alpha_{g}\simeq c_{e}/c_{g}=\frac{A+C}{B+D}.
\end{align*}

\section{$xy$ anisotropy}\label{app:xyanisotropy}
For the general set-up, the model action is given by
\begin{equation}
\begin{aligned}
S=&\int d^{3}x d\tau\;\psi^{\dagger}(\partial_{\tau}+i\tau_{z}(v_{x}\sigma_{x}\notag\partial_{x}+v_{y}\sigma_{y}\partial_{y}+v_{z}\sigma_{z}\partial_{z}))\psi\\
&+\int d^{3}x d\tau\;\psi^{\dagger}(ie\varphi+g\phi M)\psi\notag\\
&+\int d^{3}xd\tau\;\frac{1}{2}\left[ (\partial_{x}\varphi)^{2}+ R_{c,yx}^{2}(\partial_{y}\varphi)^{2}+R_{c,zx}^{2}(\partial_{z}\varphi)^{2}\right]\notag\\
&+\int d^{3}xd\tau\;\frac{1}{2}\left[\frac{(\partial_{\tau}\phi)^{2}}{u_{x}^{2}}+(\partial_{x}\phi)^{2}+ \left(\frac{u_{y}}{u_{x}}\right)(\partial_{y}\phi)^{2}+\left(\frac{u_{z}}{u_{x}}\right)^{2}(\partial_{z}\phi)^{2} \right].
\end{aligned}\end{equation}
In main text, we mentioned that $u_{x}/u_{y}=v_{x}/v_{y}=R_{c,y}=1$. In this section, we will prove this. 

Let $R_{f,yx}:=v_{y}/v_{x}$ and $R_{o,yx}:=u_{y}/u_{x}$. By using the momentum-shell RG procedure, we obtain the flow equations of $R_{f,yx}$, $R_{o,yz}$, and $R_{c,yz}$ as follows:
\begin{align*}
\frac{dR_{f,yx}}{d\ell}%=&R_{f,yx}(-\Sigma_{ky}^{\varphi}-\Sigma_{ky}^{\phi}+\Sigma_{kx}^{\varphi}+\Sigma_{kx}^{\phi})\\
=&R_{f,yx}\left[	\frac{\alpha_{e}}{\pi R_{f,yx}}\left(	h_{y}-h_{x}	\right)+\frac{\alpha_{g}}{\pi R_{f,yx}}R_{x}^{2}\left(	f_{y}-f_{x}	\right)	\right]\\
%%%%
\frac{dR_{b,yx}}{d\ell}%=&\frac{R_{b,yx}}{2}(\Pi_{kx}^{\phi}-\Pi_{ky}^{\phi})\\
%=&-\frac{2\alpha_{g}}{3\pi}\frac{R_{b,yx}}{2}\left(	\frac{1}{R_{f,yx}}-\frac{R_{x}^{2}}{R_{f,yx}R_{y}^{2}}	\right)\\
%=&-\frac{\alpha_{g}}{3\pi}\frac{R_{b,yx}}{R_{f,yx}}\left(	1-\frac{R_{x}^{2}}{R_{y}^{2}}	\right)\\
%=&-\frac{\alpha_{g}}{3\pi}\frac{R_{b,yx}}{R_{f,yx}}\left(	1-\frac{R_{f,yx}^{2}}{R_{b,yx}^{2}}	\right)\\
=&-N_{f}\frac{\alpha_{g}}{3\pi}\frac{R_{f,yx}}{R_{b,yx}}\left(	\frac{R_{b,yx}^{2}}{R_{f,yx}^{2}}-1	\right),\\
%%%%
\frac{dR_{c,yx}}{d\ell}%=&\frac{R_{c,yx}}{2}(\Pi_{kx}^{\varphi}-\Pi_{ky}^{\varphi})\\
%=&-\frac{2\alpha_{e}}{3\pi}\frac{R_{c,yx}}{2}\left(	\frac{1}{R_{f,yx}}-\frac{R_{f,yx}}{R_{c,yx}^{2}}	\right)\\
%=&-\frac{\alpha_{e}}{3\pi}\frac{R_{c,yx}}{R_{f,yx}}\left(	1-\frac{R_{f,yx}^{2}}{R_{c,yx}^{2}}	\right)\\
=&-N_{f}\frac{\alpha_{e}}{3\pi}\frac{R_{f,yx}}{R_{c,yx}}\left(	\frac{R_{c,yx}^{2}}{R_{f,yx}^{2}}-1	\right),
%%%%
\end{align*}
where $R_{x,y}=u_{x,y}/v_{x,y}$, $f_{x,y}=f_{x,y}(R_{x},R_{y},R_{z})$ and $h_{x,y}=h_{x,y}=(R_{c,yx}/R_{f,yx},R_{fc}^{-1})$ defined as follows:
\begin{align*}
h_{x}(a,b)=&\frac{1}{2\pi}\iiint \frac{(1-x^{2}+y^{2}+z^{2})}{(1+x^{2}+y^{2}+z^{2})^{2}(x^{2}+a^{2}y^{2}+b^{2}z^{2})}dxdydz,\\
h_{y}(a,b)=&\frac{1}{2\pi}\iiint \frac{(1+x^{2}-y^{2}+z^{2})}{(1+x^{2}+y^{2}+z^{2})^{2}(x^{2}+a^{2}y^{2}+b^{2}z^{2})}dxdydz,\\
f_{x}(a,b)=&\frac{1}{2\pi}\iiint \frac{(1-x^{2}+y^{2}+z^{2})}{(1+x^{2}+y^{2}+z^{2})^{2}(1+a^{2}x^{2}+b^{2}y^{2}+c^{2}z^{2})}dxdydz,\\
f_{y}(a,b)=&\frac{1}{2\pi}\iiint \frac{(1+x^{2}-y^{2}+z^{2})}{(1+x^{2}+y^{2}+z^{2})^{2}(1+a^{2}x^{2}+b^{2}y^{2}+c^{2}z^{2})}dxdydz.
\end{align*}
%
%\begin{align*}
%\frac{d(R_{b,yx}/R_{f,yx})}{d\ell}=&	\frac{1}{R_{f,yx}}\frac{dR_{b,yx}}{d\ell}-\frac{R_{b,yx}}{R_{f,yx}^{2}}\frac{dR_{f,yx}}{d\ell}\\
%=&\frac{R_{f,yx}}{R_{b,yx}}\left[		-\frac{\alpha_{g}}{3\pi}\frac{1}{R_{b,yx}}\left(\frac{R_{b,yx}^{2}}{R_{f,yx}^{2}}-1\right)		\right]
%\end{align*}
%We need to check that $R_{f,yx}=R_{o,yx}=R_{c,yx}$ is fixed point. From $dR_{b,yx}/d\ell$ and $dR_{c,yx}/d\ell$, we can show that $R_{b,yx}=R_{f,yx}$ and $R_{c,yx}=R_{f,yx}$ are stable fixed point. Near that point, 

Near $a=1$, $h_{y}-h_{x}$ becomes
\begin{align*}
h_{y}(a,b)-h_{x}(a,b)=&\frac{1}{\pi}\iiint \frac{x^{2}-y^{2}}{(1+x^{2}+y^{2}+z^{2})^{2}(x^{2}+a^{2}y^{2}+b^{2}z^{2})}dxdydz\\
(\text{around }a=1)\approx&\frac{1}{\pi}\iiint \frac{x^{2}-y^{2}}{(1+x^{2}+y^{2}+z^{2})^{2}(x^{2}+(1+2\delta a)y^{2}+b^{2}z^{2})}dxdydz\\
\approx&\frac{1}{\pi}\iiint \frac{x^{2}-y^{2}}{(1+x^{2}+y^{2}+z^{2})^{2}(x^{2}+y^{2}+b^{2}z^{2})}dxdydz\\
&\quad- 2\delta a\frac{1}{\pi}\iiint\frac{y^{2}(x^{2}-y^{2})}{(1+x^{2}+y^{2}+z^{2})^{2}(x^{2}+y^{2}+b^{2}z^{2})}dxdydz\\
=&- 2\delta a\frac{1}{\pi}\iiint\frac{y^{2}(x^{2}-y^{2})}{(1+x^{2}+y^{2}+z^{2})^{2}(x^{2}+y^{2}+b^{2}z^{2})}dxdydz\\
=&- \delta a\int\frac{r^{5}}{(1+r^{2}+z^{2})^{2}(r^{2}+b^{2}z^{2})^{2}}dr,
\end{align*}
and near $b=a$, $f_{y}-f_{x}$ becomes
\begin{align*}
f_{y}(a,b,c)-f_{x}(a,b,c)=&\frac{1}{\pi}\iiint \frac{x^{2}-y^{2}}{(1+x^{2}+y^{2}+z^{2})^{2}(1+a^{2}x^{2}+b^{2}y^{2}+c^{2}z^{2})}dxdydz\\
(\text{around }b=a)\approx&\frac{1}{\pi}\iiint \frac{x^{2}-y^{2}}{(1+x^{2}+y^{2}+z^{2})^{2}(1+a^{2}(x^{2}+y^{2}+2\delta b y^{2}/a)+b^{2}z^{2})}dxdydz\\
\approx&\frac{1}{\pi}\iiint \frac{x^{2}-y^{2}}{(1+x^{2}+y^{2}+z^{2})^{2}(1+a^{2}x^{2}+a^{2}y^{2}+c^{2}z^{2})}dxdydz\\
&\quad- 2a\delta b\frac{1}{\pi}\iiint\frac{y^{2}(x^{2}-y^{2})}{(1+x^{2}+y^{2}+z^{2})^{2}(1+a^{2}x^{2}+a^{2}y^{2}+c^{2}z^{2})}dxdydz\\
=&- 2a\delta b\frac{1}{\pi}\iiint\frac{y^{2}(x^{2}-y^{2})}{(1+x^{2}+y^{2}+z^{2})^{2}(1+a^{2}x^{2}+a^{2}y^{2}+c^{2}z^{2})}dxdydz\\
=&- \delta b\int\frac{r^{5}}{(1+r^{2}+z^{2})^{2}(1+a^{2}r^{2}+b^{2}z^{2})^{2}}dr.
\end{align*}
Using these, we can find the RG equations near $u_{x}/u_{y}=v_{x}/v_{y}=R_{c,yx}=R_{0}$
\begin{align*}
\frac{d\delta R_{f,yx}}{d\ell}=&R_{0}\left[-\frac{\alpha_{g}}{\pi}C_{1}(R_{x},R_{z})R_{x}-\frac{\alpha_{e}}{\pi}C_{2}(R_{fc})R_{x}	\right]\delta R_{f,yx},\\
\frac{d\delta R_{b,yx}}{d\ell}=&-N_{f}\frac{2\alpha_{g}}{3\pi}\frac{\delta R_{b,yx}}{R_{0}^{2}},\\
\frac{d\delta R_{c,yx}}{d\ell}=&-N_{f}\frac{2\alpha_{e}}{3\pi}\frac{\delta R_{c,yx}}{R_{0}^{2}},
\end{align*}
where $\delta R_{i,yx}=R_{i,yx}-R_{0}$ ($i=f,b,c$), and
\begin{align*}
C_{1}(a,b)=&\int\frac{r^{5}}{(1+r^{2}+z^{2})^{2}(1+a^{2}r^{2}+b^{2}z^{2})^{2}},\\
C_{2}(c)=&\int\frac{r^{5}}{(1+r^{2}+z^{2})^{2}(r^{2}+z^{2}/c^{2})^{2}}.
\end{align*}
Since $C_{1}>0$ and $C_{2}>0$, $\delta R_{f,yx}$, $\delta R_{b,yx}$, and $\delta R_{c,yx}$ vanish. Therefore, the fixed point $R_{f,yx}=R_{c,yx}=R_{b,yx}$ is stable. Then, for general $R_{f,yx}=R_{c,yx}=R_{b,yx}=R_{0}$, the flow equations of the remaining dimensionless parameters are given by
\begin{align}\notag
\frac{d R_{\perp}}{d\ell}=& \frac{R_{\perp}}{R_{0}}\left[ \frac{\alpha_{g}}{\pi}\left( \frac{N_{f}}{3}(1- R_{\perp}^{2})+F_{x}( R_{\perp}, R_{z}) \right)-\frac{\alpha_{e}}{\pi}H_{x}( R_{fc}) \right],\\\notag
\frac{d R_{z}}{d\ell}=& \frac{R_{z}}{R_{0}}\left[-\frac{\alpha_{g}}{\pi}\left( \frac{ R_{\perp}^{2}}{3}N_{f}-F_{z}( R_{\perp}, R_{z}) \right)-\frac{\alpha_{e}}{\pi}H_{z}( R_{fc})\right],\\\notag
\frac{d R_{fc}}{d\ell}=& \frac{R_{fc}}{R_{0}}\left[  -\frac{\alpha_{g}}{\pi}\left(F_{z}( R_{\perp}, R_{z})-F_{x}( R_{\perp}, R_{z})\right)\right.\\\notag
&\quad\left.-\frac{\alpha_{e}}{\pi}\left( \frac{N_{f}}{3}( R_{fc}^{2}-1)+(H_{x}( R_{fc})-H_{z}( R_{fc})) \right)\right],\\\notag
\frac{d\alpha_{g}}{d\ell}=&\frac{\alpha_{g}}{R_{0}}\left[ -\frac{\alpha_{g}}{\pi}\left(\frac{2}{3}N_{f}+F_{z}( R_{\perp}, R_{z})\right)+\frac{\alpha_{e}}{\pi}H_{z}( R_{fc}) \right],\\\notag
\frac{d\alpha_{e}}{d\ell}=&\frac{\alpha_{e}}{R_{0}}\left[ -\frac{\alpha_{e}}{\pi}\left(\frac{2}{3}N_{f}+H_{z}( R_{fc})\right)+\frac{\alpha_{g}}{\pi}F_{z}( R_{\perp}, R_{z}) \right],\\
\frac{d\tilde{\lambda}}{d\ell}=&\frac{\tilde{\lambda}}{R_{0}}\left[-\frac{3\tilde{\lambda}}{16\pi^{2}}-\frac{N_{f}}{3\pi}\alpha_{g}(2+ R_{\perp}^{2})\right].
\end{align}
In comparison to the flow equations in main text, the difference is only overall factor of $R_{0}^{-1}$. However, it does not affect on the result of our low energy analysis. Therefore, we can set $R_{0}=1$ because the result is not changed.

\section{Details of loop functions}\label{app:loopfunction}
The loop functions $F_{x}$, $F_{z}$, $H_{x}$, and $H_{z}$ are defined by
\begin{align}
\notag F_{x}( a, b)\equiv&\frac{ a^{2}}{\pi}\int_{0}^{\infty}dr\int_{-\infty}^{\infty}dy\;\frac{r(r^{2}-2)}{(1+r^{2}+y^{2})^{2}(1+ a^{2}r^{2}+ b^{2}y^{2})}\\
=&\frac{ a^{2}}{2}\left[ \frac{ b( a^{2}-1)+3( a^{2}- b^{2})}{(1+ b)( a^{2}- b^{2})( a^{2}-1)} -\frac{((2 a^{2}+1) a^{2}-( a^{2}+2) b^{2})}{( a^{2}-1)^{3/2}( a^{2}- b^{2})^{3/2}}\tanh^{-1}\left( \frac{\sqrt{ a^{2}-1}\sqrt{ a^{2}- b^{2}}}{ a^{2}+b} \right)\right],\label{eqA:Fx}\\
\notag F_{z}( a, b)\equiv&\frac{ a^{2}}{\pi}\int_{0}^{\infty}dr\int_{-\infty}^{\infty}dy\;\frac{2r^{3}}{(1+r^{2}+y^{2})^{2}(1+ a^{2}r^{2}+ b^{2}y^{2})}\\
=& a^{2}\left[ \frac{ a^{2}- b}{( a^{2}-1)( a^{2}- b^{2})}-\frac{( a^{2}( b^{2}+1)-2 b^{2})}{( a^{2}-1)^{3/2}( a^{2}- b^{2})^{3/2}}\tanh^{-1}\left( \frac{\sqrt{ a^{2}-1}\sqrt{ a^{2}- b^{2}}}{ a^{2}+b} \right) \right],\label{eqA:Fz}\\
\notag H_{x}( c)\equiv&\frac{1}{\pi}\int_{0}^{\infty}dr\int_{-\infty}^{\infty}dy\;\frac{r(1+y^{2})}{(1+r^{2}+y^{2})^{2}(r^{2}+y^{2}/ c^{2})}\\
=&\frac{ c}{2}\left[ \frac{ c}{1- c^{2}}+\frac{(2 c^{2}-1)}{( c^{2}-1)^{3/2}}\tanh^{-1}\left( \frac{\sqrt{ c^{2}-1}}{ c} \right) \right],\label{eqA:Hx}\\
\notag H_{z}( c)\equiv&\frac{1}{\pi}\int_{0}^{\infty}dr\int_{-\infty}^{\infty}dy\;\frac{r(1+r^{2}-y^{2})}{(1+r^{2}+y^{2})^{2}(r^{2}+y^{2}/ c^{2})}\\
=& c\left[ \frac{ c}{ c^{2}-1}-\frac{1}{( c^{2}-1)^{3/2}}\tanh^{-1}\left( \frac{\sqrt{ c^{2}-1}}{ c}\right) \right],\label{eqA:Hz}
\end{align}
where $F_{x}$ and $F_{z}$ come from the order parameter-fermion loop integrals, and $H_{x}$ and $H_{z}$ come from coulomb interaction-fermion loop integrals.

From the analytic expressions of the loop functions, we can know that 
\begin{align*}
H_{x}(c)=&F_{z}(a=c,b=0)-F_{x}(a=c,b=0),\\
H_{z}(c)=&F_{z}(a=c,b=0).
\end{align*}
Therefore, by investigating $F_{x}$ and $F_{z}$, we can know about $H_{x}$ and $H_{z}$.

\begin{figure}
\centering
\subfigure[]{
\includegraphics[width=0.45\linewidth]{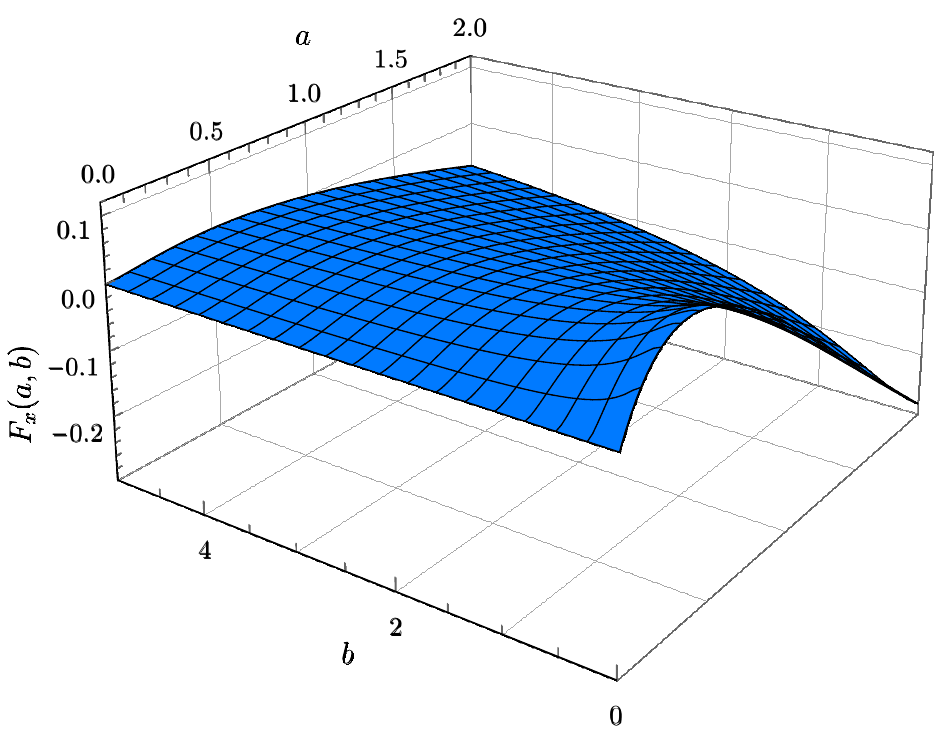}\label{fig:fx3d}}\subfigure[]{
\includegraphics[width=0.45\linewidth]{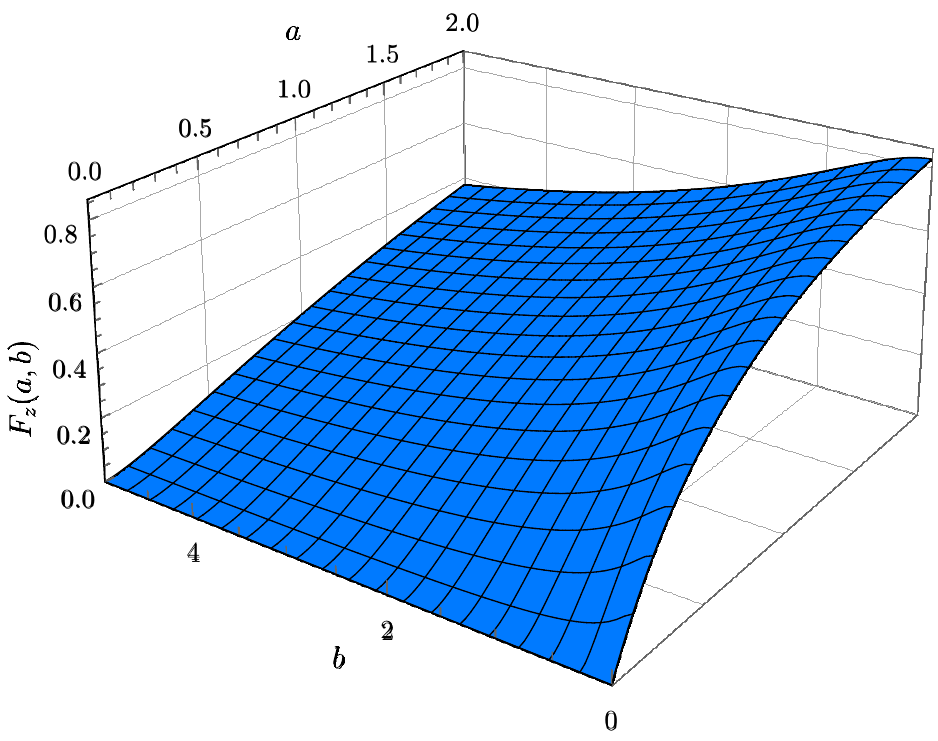}\label{fig:fz3d}}
\subfigure[]{
\includegraphics[width=0.45\linewidth]{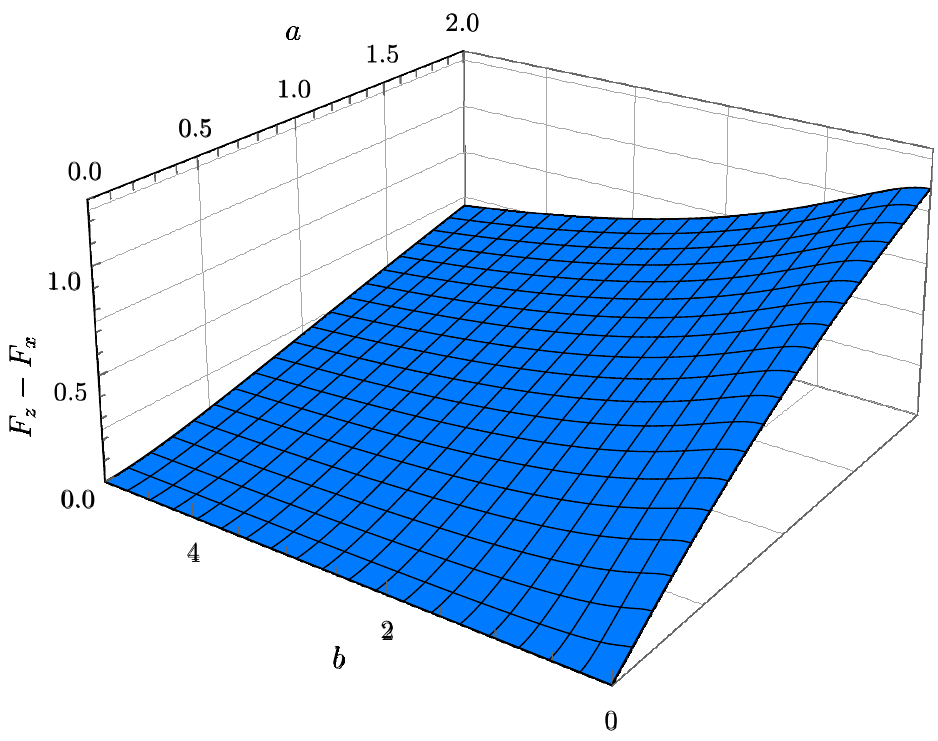}\label{fig:fzfx3d}
}\subfigure[]{
\includegraphics{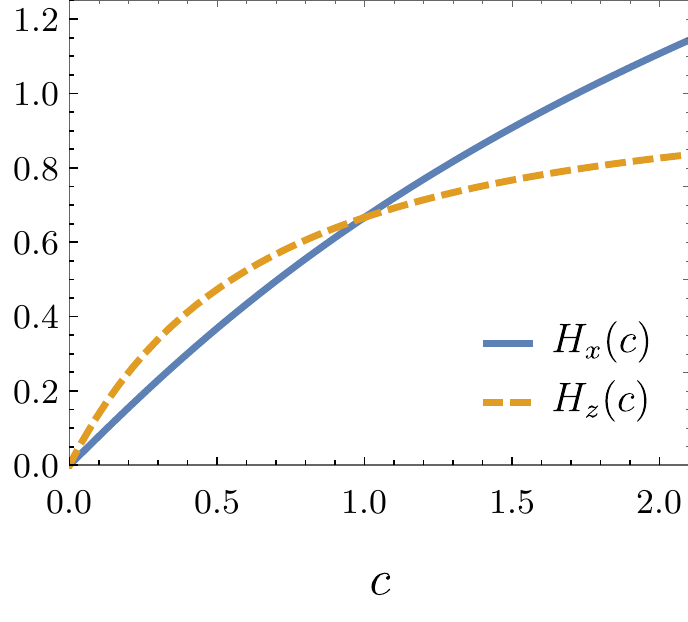}\label{fig:hxhz}
}
\caption{ The loop functions $F_{x}$, $F_{z}$, $H_{x}$ and $H_{z}$. (a) The functions $F_{x}(a,b)$ in terms of $a$ and $b$. $F_{x}(a,b)=0$ at $a=1$, $0\leq F_{x}(a,b)<0.113$ for $a<1$ and $F_{x}(a,b)<0$ for $a>1$. (b) The functions $F_{z}(a,b)$ in terms of $a$ and $b$. $F_{z}$ is positive semi-definite for all $a$ and $b$. (c) The profile of $F_{z}-F_{x}$. It is positive semi-definite for all $a$ and $b$. (d) $H_{x}(c)$ and $H_{z}(c)$ in terms of $c$. The blue solid (orange dashed) lined is for $H_{x}(c)$ ($H_{z}(c)$). There are monotonic increasing functions and have the same value $2/3$ at $c=1$.}
\end{figure}

For $F_{x}(a,b)$, it has the relation in terms of $a$,
\begin{equation}\label{eq:cond1}
\begin{split}
F_{x}(a,b)\geq0,&\quad\text{for }a<1,\\
F_{x}(a,b)<0,&\quad\text{for }a>1,
\end{split}
\end{equation}
and $F_{x}(a,b)=0$ at $a=1$ (see Fig.\ref{fig:fx3d}).

For $F_{z}$, it is positive semi-definite for all $a$ and $b$, $F_{z}(a,b)\geq0$ (see Fig.\ref{fig:fz3d}). At $a=1$, it has the value
\begin{align*}
F_{z}(a=1,b)=\frac{2(1+2b)}{3(1+b)^{2}}\leq\frac{2}{3}.
\end{align*}
And for all $a$ and $b$, $F_{z}(a,b)\geq F_{x}(a,b)$ (see Fig.\ref{fig:fzfx3d}).

As mentioned before, from the properties of $F_{x}$ and $F_{z}$, we can deduce the properties of $H_{x}$ and $H_{z}$. For $H_{x}(c)$ and $H_{z}(c)$, they are monotonic increasing function in terms of $c$ (see Fig,\ref{fig:hxhz}). As you can see, they have the same value $2/3$ at $c=1$ because of $F_{x}(1,0)=0$ and $F_{z}(1,0)=2/3$. Also, So, $H_{x}(c)<H_{z}(c)$ for $c<1$, but $H_{x}(c)>H_{z}(c)$ for $c>1$.

\section{$N_{f}$ Dirac fermions with long-range Coulomb interaction}\label{app:coulombrg}
Let us consider the situation which we have $N_{f}$ Dirac fermions with long-range Coulomb interaction. This can be obtained by ignoring $\alpha_{g}$ in Eq.\ref{eq:beta}. Here, we consider the dimensionless parameters, $R_{fc}$ and $\alpha_{e}$. The RG flow equations for dimension parameters are
\begin{align*}
\frac{dR_{fc}}{d\ell}=&-\frac{\alpha_{e}}{\pi}R_{fc}\left[	\frac{N_{f}}{3}(R_{fc}^{2}-1)+(H_{x}(R_{fc})-H_{z}(R_{fc}))	\right],\\
\frac{d\alpha_{e}}{d\ell}=&-\frac{\alpha_{e}^{2}}{\pi}\left[	\frac{2}{3}N_{f}+H_{z}(R_{fc})	\right],
\end{align*}
and the flow equations for the anisotropy constants are
\begin{align*}
\frac{d\gamma_{f}}{d\ell}=&\frac{\alpha_{e}}{\pi}\gamma_{f}\left[	H_{z}(R_{fc})-H_{x}(R_{fc})	\right],\\
\frac{d\gamma_{c}}{d\ell}=&-\frac{N_{f}\alpha_{e}}{3\pi}\gamma_{c}(1-R_{fc}^{2}).
\end{align*}
For $R_{fc}$, its fixed point value is $R_{fc}^{*}=1$. To check this, expanding near $R_{fc}^{*}=1$, $R_{fc}\approx1+\delta R_{fc}$,
\begin{align*}
\frac{d\delta R_{fc}}{d\ell}=&-\frac{\alpha_{e}}{\pi}\left[	\frac{2}{3}N_{f}	+\frac{4}{15}\right]\delta R_{fc},
\end{align*}
so it vanishes. For $\alpha_{e}$, its flow equation is always negative, so it also vanishes. Let us consider the anisotropy constants of fermion and Coulomb interaction. The fixed point value $R_{fc}^{*}=1$ means that $\gamma_{f}^{*}=\gamma_{c}^{*}=\gamma_{0}$ where $\gamma_{0}$ is constant between $\gamma_{f,0}$ and $\gamma_{c,0}$ (subscript 0 stands for initial value). Near $\gamma_{0}$, $\gamma_{f,c}\approx \gamma_{0}(1+\delta\gamma_{f,c})$, the flow equations for the anisotropy constants are
\begin{align*}
\frac{d \delta\gamma_{f}}{d\ell}=&-\frac{4\alpha_{e}}{15\pi}\delta\gamma_{f},\\
\frac{d \delta\gamma_{c}}{d\ell}=&-\frac{2N_{f}\alpha_{e}}{3\pi}\delta\gamma_{c}.
\end{align*}
Therefore, $\delta \gamma_{f,c}\rightarrow0$.\\
Let us assume that $\gamma_{0}=1$, i.e., isotropic case. Then, the RG flow equations for $v$ and $\alpha_{e}$ are
\begin{align*}
\frac{dv}{d\ell}=&\frac{2\alpha_{e}}{3\pi}v,\\
\frac{d\alpha_{e}}{d\ell}=&-\frac{2\alpha_{e}^{2}}{3\pi}(N_{f}+1).
\end{align*}
This is usual result for $N_{f}$ Dirac fermions with long-range Coulomb interaction.

\section{Renormalization of order parameter mass $r$}\label{app:ordermass}
\begin{figure}[h]
\centering
\subfigure[]{
\includegraphics{fig1d.pdf}\label{fig:mass1}}
\subfigure[]{
\includegraphics{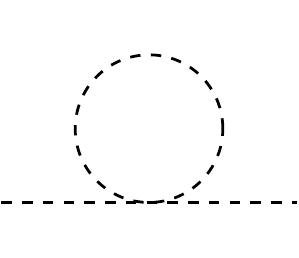}\label{fig:mass2}}
\caption{Feynman diagrams for order parameter mass at the one-loop order. The line with arrowhead, dashed line, and wavy line stand for the fermion and the order parameter, respectively.}\label{fig:ordermass}
\end{figure}

The renormalization of order parameter mass is
\begin{align*}
\delta_{r}=&-g^{2}\int_{\Omega,\bm{q}}\frac{\Omega^{2}+v_{\perp}^{2}q_{\perp}^{2}-v_{z}^{2}q_{z}^{2}}{(\Omega^{2}+v_{\perp}^{2}q_{\perp}^{2}+v_{z}^{2}q_{z}^{2})^{2}}+2\frac{\lambda}{4u_{\perp}}\int_{\Omega,\bm{q}}\frac{1}{\Omega^{2}/u_{\perp}^{2}+q_{\perp}^{2}+(u_{z}/u_{\perp})^{2}q_{z}^{2}+r^{2}/u_{\perp}^{2}}\\
=&\frac{1}{u_{\perp}^{2}}\frac{\lambda}{u_{z}/u_{\perp}}\frac{\Lambda^{2}\ell}{16\pi^{4}}\int d^{3}q\frac{1}{1+\tilde{r}+q^{2}}\\
\simeq&-\frac{1}{u_{\perp}^{2}}\frac{\lambda}{u_{z}/u_{\perp}}\frac{\Lambda^{2}\ell}{16\pi^{4}}2\pi^{2}\sqrt{1+\tilde{r}}\\
=&-\frac{1}{u_{\perp}^{2}}\frac{\tilde{\lambda}\Lambda^{2}\ell}{8\pi^{2}}\sqrt{1+\tilde{r}}.
\end{align*}
where $\tilde{r}=r/\Lambda^{2}$ and we take only logarithmic divergence in third line. In the first line, the first and second terms come from the fermion (Fig.\ref{fig:mass1}) and order parameter loop (Fig.\ref{fig:mass2}), respectively, and the fermion loop contribution vanishes. %There are the linear divergence coming from the order parameter loop, and it is absorbed in bare order parameter mass. 
Then, the RG equation for $r$ is
\begin{align*}
\frac{dr}{d\ell}=&2r-\frac{\tilde{\lambda}\Lambda^{2}}{8\pi^{2}}\sqrt{1+\tilde{r}}
\end{align*}
At the leading order, the solution of above RG equation at QCP is obtained by using the fixed-point value in the main text ($\tilde{\lambda}^{*}=0$ for all the cases), so we obtain $r_{c}\approx0$. So, the setting $r=0$ will not affect the RG analysis on the QCP and we can set $r=0$ in the RG analysis to explore QCP.

\end{document}